\documentclass[10pt]{article}

\usepackage{scicite}

\usepackage{times}

\usepackage{graphicx}

\usepackage[dvipsnames]{xcolor}

\usepackage{indentfirst}

\usepackage{multicol}

\usepackage{float}

\usepackage{bm}

\usepackage{amsmath}

\usepackage{caption}
\usepackage{setspace}
\usepackage{titling}
\usepackage[labelfont=bf]{caption}

\newenvironment{sciabstract}{%
\begin{quote}}
{\end{quote}}

\newcounter{lastnote}

\title{Filming Enhanced Ionization in an Ultrafast Triatomic Slingshot}

\author
{A.J. Howard,$^{1,2}$ M. Britton,$^{2,3}$ Z.L. Streeter,$^{4,5}$ C. Cheng,$^{6}$ R. Forbes,$^{2,7}$ \\
J.L. Reynolds,$^{1}$ F. Allum,$^{2,7}$ G.A. McCracken,$^{1,2}$ I. Gabalski,$^{1,2}$  R.R. Lucchese$^{5}$, \\
C.W. McCurdy,$^{4,5}$ T. Weinacht,$^{6}$ and P.H. Bucksbaum$^{1,2,3,7,\ast}$\\
\\
\normalsize{$^{1}$Department of Applied Physics, Stanford University,}\\ \normalsize{Stanford, California 94305, USA}\\
\normalsize{$^{2}$Stanford PULSE Institute, SLAC National Accelerator,}\\ \normalsize{Laboratory, 2575 Sand Hill Road, Menlo Park, California 94025, USA}\\
\normalsize{$^{3}$Department of Physics, Stanford University,}\\
\normalsize{Stanford, California 94305, USA}\\
\normalsize{$^{4}$Department of Chemistry, University of California, Davis,}\\
\normalsize{Davis, California 95616, USA}\\
\normalsize{$^{5}$Chemical Sciences Division, Lawrence Berkeley National Laboratory,}\\
\normalsize{Berkeley, California 94720, USA}\\
\normalsize{$^{6}$Department of Physics and Astronomy, Stony Brook University,}
\normalsize{Stony Brook, New York 11794, USA}\\
\normalsize{$^{7}$Linac Coherent Light Source, SLAC National Accelerator Laboratory,}\\
\normalsize{Menlo Park, California 94025, USA}\\
\normalsize{$^\ast$To whom correspondence should be addressed; E-mail:  phbuck@stanford.edu.}
}
\date{\vspace{-0.7cm}}

\begin{document} 


\baselineskip10pt

\maketitle 

\begin{sciabstract}
Filming atomic motion within molecules is an active pursuit of molecular physics and quantum chemistry.
A promising method is laser-induced Coulomb Explosion Imaging (CEI) where a laser pulse rapidly ionizes many electrons from a molecule, causing the remaining ions to undergo Coulomb repulsion.
The ion momenta are used to reconstruct the molecular geometry which is tracked over time (i.e. filmed) by ionizing at an adjustable delay with respect to the start of interatomic motion.
Results are distorted, however, by ultrafast motion during the ionizing pulse.
We studied this effect in water and filmed the rapid ``slingshot'' motion that enhances ionization and distorts CEI results.
Our investigation uncovered both the geometry and mechanism of the enhancement which may inform CEI experiments in many other polyatomic molecules.
\end{sciabstract}


\noindent

\begin{multicols}{2}
\noindent \fontfamily{ptm}\selectfont \textbf{INTRODUCTION} \\
\indent \fontfamily{ptm}\selectfont Hydrogen atoms within molecules can move extremely rapidly in response to the sudden intramolecular forces introduced by ionization or photoexcitation.
These atoms can accelerate to traverse Angstrom-scale distances in just tens of femtoseconds.
Such ultrafast motion may mediate many biologically important light-matter interactions, including photosynthesis, photochemical damage mitigation in DNA, and vision \cite{brixner_two-dimensional_2005,schwalb_base_2008,prokhorenko_coherent_2006}.
Resolving the femtosecond-scale motion of these light atomic species is therefore central to the fields of molecular physics and quantum chemistry, and motivates continued efforts to film atomic-scale ``molecular movies'' \cite{dwyer_femtosecond_2006,barty_molecular_2013,ivanov_concluding_2021}. \\
\indent Many methods exist to record ultrafast motion in molecules including diffractive imaging techniques, such as ultrafast electron diffraction \cite{yang_imaging_2018,wolf_photochemical_2019,yang_simultaneous_2020}, laser-induced electron diffraction \cite{meckel_laser-induced_2008,wolter_ultrafast_2016}, and hard x-ray diffraction \cite{minitti_imaging_2015,kim_direct_2015,glownia_self-referenced_2016};
spectroscopic techniques such as high harmonic generation \cite{li_time-resolved_2008,worner_conical_2011,he_monitoring_2018}; and momentum imaging techniques such as Coulomb Explosion Imaging (CEI) \cite{pitzer_direct_2013,kunitski_observation_2015,fehre_enantioselective_2019,endo_capturing_2020,erk_imaging_2014,liekhus-schmaltz_ultrafast_2015,zeller_imaging_2016,rudenko_femtosecond_2017,boll_x-ray_2022,herwig_imaging_2013,reedy_dissociation_2018,severt_step-by-step_2022}.
Diffractive techniques, however, lose sensitivity with decreasing atomic mass and spectroscopic techniques typically rely on indirect observables of atomic motion such as electronic structure.
Among the methods listed, only CEI probes the direct momenta of all atoms, irrespective of mass, within a molecule \cite{yamanouchi_visualizing_2012,liekhus-schmaltz_ultrafast_2015}.\\
\indent CEI deduces the positions of atoms within molecules by stripping away binding electrons and measuring the momenta of fragments produced in the resulting Coulomb repulsion of the ions.
The binding electrons can be removed using thin foils \cite{vager_coulomb_1989,herwig_imaging_2013}, intense infrared (IR) lasers \cite{stapelfeldt_wave_1995,pitzer_direct_2013,kunitski_observation_2015,fehre_enantioselective_2019,endo_capturing_2020} or ultrafast x-ray pulses \cite{erk_imaging_2014,liekhus-schmaltz_ultrafast_2015,zeller_imaging_2016,rudenko_femtosecond_2017,fukuzawa_real-time_2019,boll_x-ray_2022}.
Intense IR lasers are especially attractive for time-resolved CEI; intramolecular dynamics can be initiated with a few-femtosecond table-top IR laser pulse and probed with sub-femtosecond temporal jitter using another pulse derived from the same source \cite{rudenko_real-time_2006,bocharova_time-resolved_2011}. \\
\begin{figure*}[t]
\centering
\includegraphics[width=19cm, trim={0cm 0cm 0cm 0cm}]{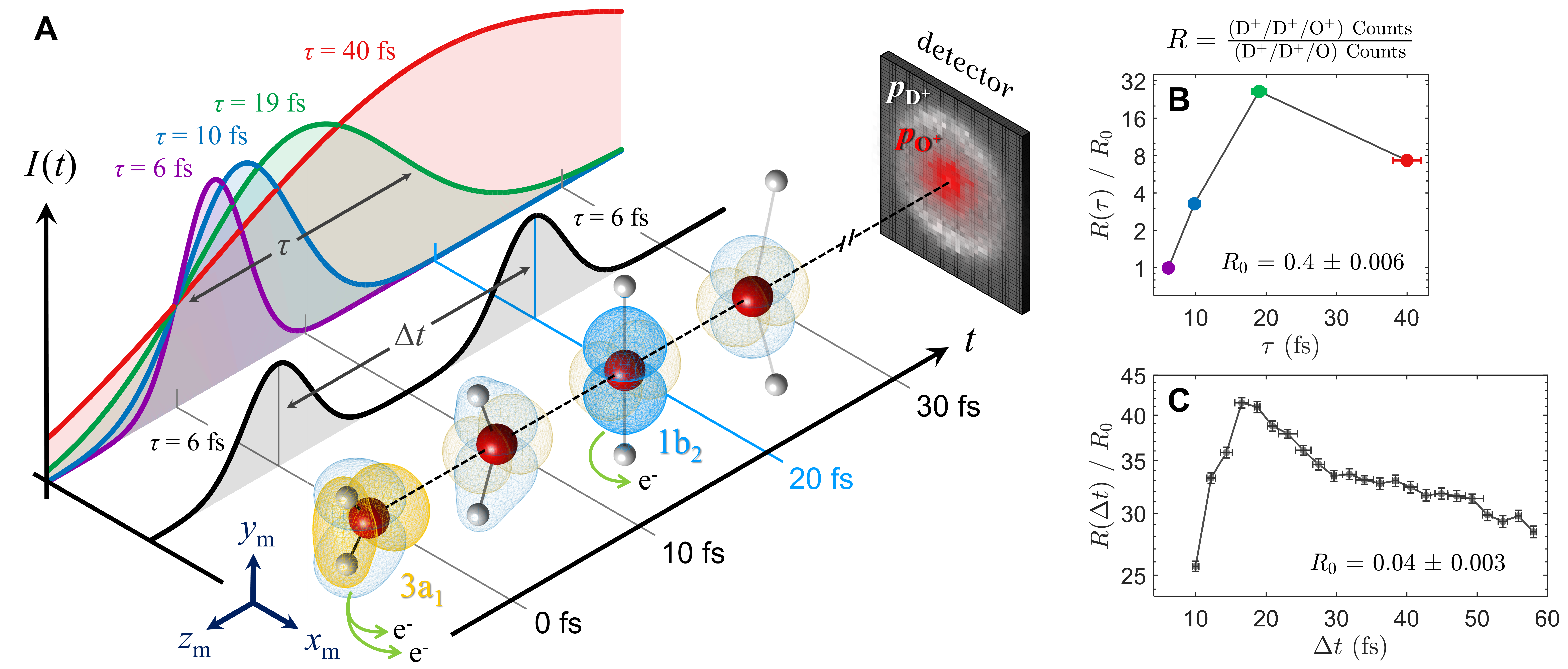}
\label{fig:Schematic}
\vspace{-0.2cm}
\caption{ \textbf{Strong-field enhanced ionization forming D$_2$O$^{3+}$.} (\textbf{A}) A schematic of the experiment.
On the left, the temporal intensity profiles, $I$($t$), are displayed for single pulses of variable duration, $\tau$, (in color) and pulse pairs with variable interpulse delay, $\Delta t$, (in black).
Ultrafast nuclear dynamics are initiated by the ionization of two electrons from neutral D$_2$O at $t$~=~0.
Upon reaching a critical geometry at $t$~=~20~fs, Enhanced Ionization (EI) occurs to facilitate the ionization of a third electron.
Following formation of D$_2$O$^{3+}$, three molecular fragments (D$^{+}$/D$^{+}$/O$^{+}$) are produced and mapped to a detector screen, each imprinting a three-dimensional momentum distribution ($p_\mathrm{D^{+}}$ and $p_\mathrm{O^{+}}$) dependent on the molecular geometry prior to dissociation.
Two molecular orbitals, 3a$_1$ and 1b$_2$, are highlighted as the molecular geometry distorts.
(\textbf{B}) The measured ratio ($R$) of triply-charged (D$^{+}$/D$^{+}$/O$^{+}$) to doubly-charged (D$^{+}$/D$^{+}$/O) three-body dissociations plotted logarithmically as a function of pulse duration (where I$_0$~=~2$\times$10$^{15}~$W/cm$^2$) and scaled to the ratio for a 6-fs pulse ($R_0$~=~0.4~$\pm$~0.006). 
(\textbf{C}) The same ratio as in (B) plotted as a function of interpulse delay (where I$_0$~=~1$\times$10$^{15}~$W/cm$^2$) and scaled to the ratio for a single 6-fs pulse ($R_0$~=~0.04~$\pm$~0.003).
In both panels (B) and (C), EI manifests as $R/R_0$~$>$~1.}
\end{figure*}
\indent One of the greatest limitations of laser-induced CEI is that multiple ionization typically occurs sequentially; as a result, intermediate charge states can drive nuclear dynamics prior to Coulomb explosion \cite{legare_laser_2005,legare_laser_2006}.
The ultimate dissociation pathway of the fragments is therefore rarely determined by idealized Coulomb repulsion between point-like ions.
This problem is greatly exacerbated by a strong-field phenomenon known as Enhanced Ionization (EI), where a ``critical'' spacing among the constituent atoms of a molecule increases the ionization yield \cite{seideman_role_1995,zuo_charge-resonance-enhanced_1995,bocharova_charge_2011,liu_charge_2015,wu_probing_2012}, distorting the momentum distribution observed via CEI to favor the critical geometries that undergo EI. \\
\indent Strong-field distortions of molecular dissociation dynamics were first studied extensively in diatomic molecules ~\cite{ibrahim_h_2018,bucksbaum_softening_1990,zuo_charge-resonance-enhanced_1995,seideman_role_1995} and have been more recently studied in triatomic molecules~\cite{legare_laser_2005,legare_laser_2006,bocharova_charge_2011,zhao_strong-field-induced_2019} such as water \cite{liu_charge_2015,mccracken_geometric_2017,mccracken_ionization_2020,cheng_momentum-resolved_2020,howard_strong-field_2021,cheng_strong-field_2021,allum_multi-particle_2021}.
Diatomics have long been known to undergo EI en route to dication formation: strong-field ionization of the first electron drives nuclear motion toward a geometry that facilitates ionization of the second electron.
In this geometry the two atoms are always stretched to a critical distance and aligned with the polarization axis of the laser \cite{trump_multiphoton_1999,trump_strong-field_1999,ergler_time-resolved_2005,ben-itzhak_elusive_2008,xu_experimental_2015}.
If both electrons are removed within the same pulse, it has been found that decreasing the pulse duration generally reduces the fraction of molecules that undergo EI \cite{legare_time-resolved_2003,legare_laser_2005,legare_laser_2006}.
A similar phenomenon exists in multiply-charged linear triatomics \cite{bocharova_charge_2011,matsuda_time-resolved_2014}, and has recently been studied both experimentally \cite{liu_charge_2015} and theoretically \cite{koh_ionization_2020} in triply charged water, the bent triatomic considered in this work. \\
\indent Here, we used D$_2$O as a model molecular system to film an EI process that proceeds within just 20 fs.
To do so, we first verified that the formation of D$_2$O$^{3+}$ via strong-field multiple ionization yields CEI results with clear distortions indicative of EI.
We characterized how the severity of these distortions greatly depends on the ionizing pulse duration.
We then demonstrated that the conditions for EI can be reproduced by launching rapid ``slingshot'' motion in D$_2$O$^{2+}$.
We modeled this motion using \it{ab initio}\rm{} theory, and the correspondence between theory and experiment allowed the direct retrieval of the time-resolved molecular geometry.
The resulting molecular movie revealed the critical geometry at which EI occurs and unveiled the underlying mechanism that induces the enhancement.
This improved understanding of EI can not only aid the analysis of future CEI data, but can also be employed to highlight particular features of polyatomic motion in future atomic-scale molecular movies.
\\
\\
\\
\begin{figure*}[t]
\centering
\includegraphics[width=19cm, trim={0cm 0cm 0cm 0cm}]{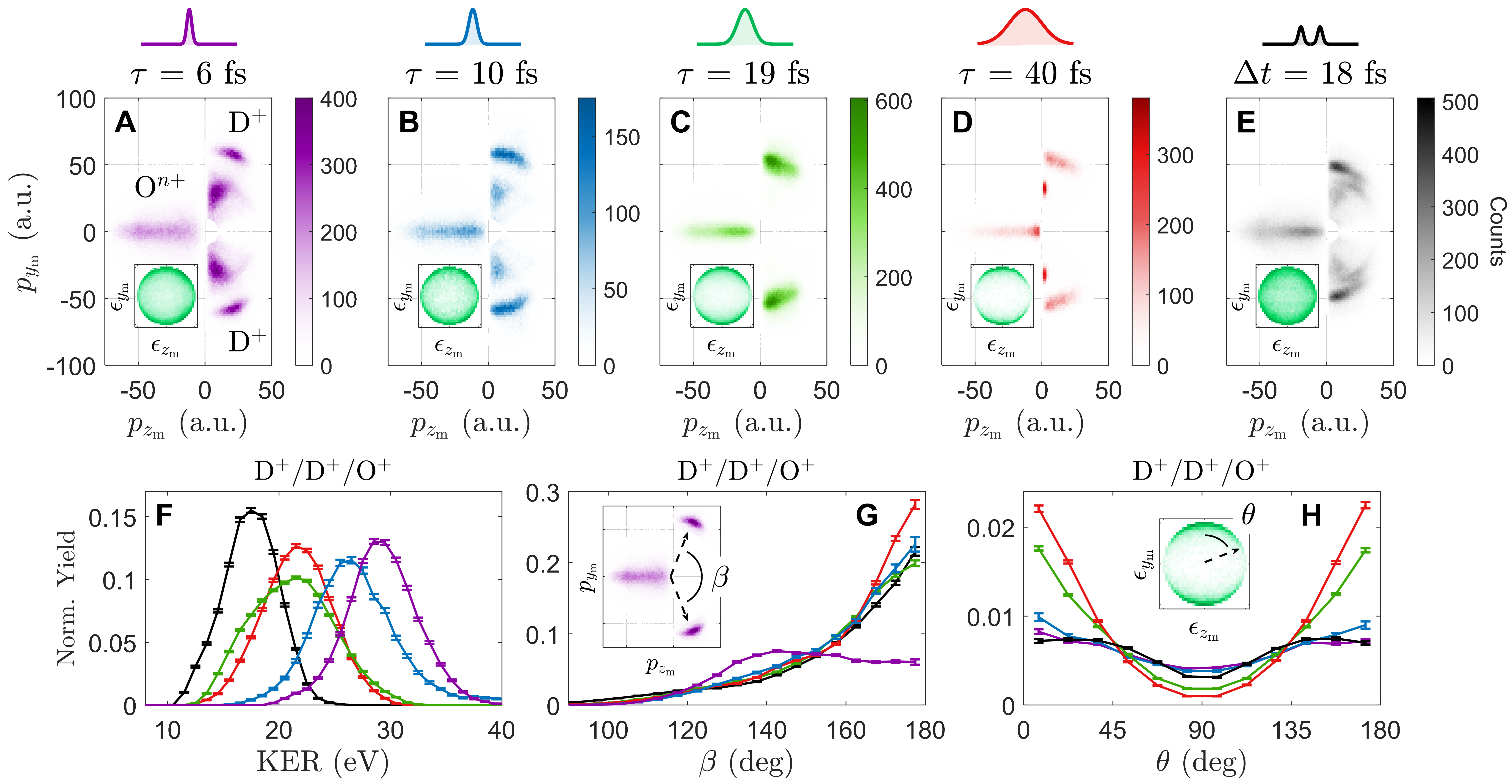}
\label{fig:Newton}
\vspace{-0.2cm}
\caption{ \textbf{Pulse-shape dependence of D$_2$O$^{2+}$~$\&$~D$_2$O$^{3+}$ fragment momenta.} (\textbf{A}-\textbf{E}) The molecular-frame momentum distribution (in $p_{z_\mathrm{m}}$ and $p_{y_\mathrm{m}}$) of the three fragments (D$^{+}$/D$^{+}$/O$^{n+}$ where $n$~=~0,1) present in all D$^{+}$/D$^{+}$ coincidences.
In each plot, the D$^+$ momentum distributions appear in the upper and lower right quadrants ($p_{z_\mathrm{m}}>0$) while the O$^{n+}$ momentum distribution appears on the left ($p_{z_\mathrm{m}}<0$).
Panels~\mbox{(A-D)} correspond to ionization via single pulses with durations ranging from 6 to 40~fs.
Panel~(E) corresponds to ionization via pulse pairs with an interpulse delay of 18~fs.
The inset in panels (A-E) displays a crushed 2-dimensional projection (in $\epsilon_{z_{\mathrm{m}}}$ and $\epsilon_{y_{\mathrm{m}}}$) of the normalized 3-dimensional polarization vector in the molecular frame. (\textbf{F}-\textbf{H}) One-dimensional distributions of the total kinetic energy release (KER), momentum-frame bend-angle, \mbox{$\beta=\mathrm{arccos}(\hat{p}_\mathrm{D_{(1)}^{+}} \cdot \hat{p}_\mathrm{D_{(2)}^{+}})$}, and molecular-frame alignment angle, $\theta=\mathrm{arccos}(\hat{y}_{\mathrm{m}} \cdot \hat{\epsilon})$, for each D$^{+}$/D$^{+}$/O$^{+}$ distribution displayed in panels (A-E).}
\end{figure*}
\noindent \fontfamily{ptm}\selectfont \textbf{RESULTS} \\
\noindent \fontfamily{ptm} \textbf{Eliciting EI in Water} \\
\indent \fontfamily{ptm}\selectfont Two experimental schemes were used to induce EI in water, both of which are depicted in Fig.~1A.
In the first scheme, 800-nm pulses of variable duration ($\tau$~=~6,~10,~19,~and~40~fs) but constant peak intensity (I$_0$~=~2$\times$10$^{15}$~W/cm$^2$) ionize neutral D$_2$O to form D$_2$O$^{3+}$.
Here, the intermediate charge states, D$_2$O$^{+}$ and D$_2$O$^{2+}$, undergo field-assisted dynamics within the envelope of a single pulse.
In the second scheme, a 6-fs 750-nm pulse (I$_0$~=~1$\times$10$^{15}$~W/cm$^2$) doubly ionizes the neutral molecule to create the dication D$_2$O$^{2+}$ before an identical cross-polarized pulse follows at an adjustable delay ($\Delta t$~=~10 to 110~fs) to form the trication D$_2$O$^{3+}$.
Here, the dication undergoes field-free dynamics in the time between pulses.
Following trication formation, the molecule rapidly Coulomb exploded into three bodies (D$^+$/D$^+$/O$^+$), each of whose three-dimensional momenta were captured in coincidence by a high-resolution position- and time-sensitive detector \cite{jagutzki_multiple_2002}. (See Materials and Methods for further details on the experimental design.)
\\
\indent Both long pulses and pulse pairs set to a particular interpulse delay were found to more efficiently strip three electrons from D$_2$O than single short pulses at the same intensity (Fig. 1B and 1C.)
For single pulses, we found that the ratio ($R$) of triply- (D$^{+}$/D$^{+}$/O$^{+}$) to doubly-charged (D$^{+}$/D$^{+}$/O) three-body dissociations undergoes a 27-fold increase as the pulse duration lengthens from 6 to 19~fs (Fig.~1B).
Using pulse pairs, a lower peak intensity was chosen to purposely highlight any enhancement, and consequently $R$ increased nearly 42 times when the interpulse delay was set to 18~fs as compared to a single pulse at the same intensity (Fig.~1C).
We now examine the CEI observables in the one-pulse and two-pulse data to find the dynamics responsible for each enhancement.\\
\indent The fragment momenta captured following ionization with single pulses are strongly suggestive of ultrafast nuclear motion.
These momenta were transformed to the molecular frame defined (by coordinates $x_\mathrm{m}$, $y_\mathrm{m}$, and $z_\mathrm{m}$) in the lower left corner of Fig.~1A
(see Materials and Methods for a mathematical description of this transformation).
\mbox{Figs.~2A--2D} plot the molecular-frame momenta of all three fragments present in every D$^+$/D$^+$ coincidence following ionization at each of the four pulse durations studied.
In Fig.~2A, for example, the higher-momentum D$^+$ cluster results from triple ionization (D$^+$/D$^+$/O$^+$), and the lower-momentum D$^+$ cluster results from double ionization (D$^+$/D$^+$/O) \cite{howard_strong-field_2021,cheng_strong-field_2021,allum_multi-particle_2021}.
The ratio of these two fragmentation pathways for the four pulse durations is shown in Fig.~1B.
(Fig.~2D reveals why this ratio ultimately decreases as the pulse duration is increased from 19 to 40 fs: a narrowly distributed D$^+$/D$^+$/O feature appears when $\tau$~=~40 fs which we attribute to the relatively slow unbending of D$_2$O$^{+}$ \cite{cheng_strong-field_2021} facilitating second ionization and subsequent three-body dissociation.)
As this ratio changes, the momentum distributions within the D$^+$/D$^+$/O$^+$ channel also change due to particular geometrical distortions: stretching of the OD bond lengths ($r_\mathrm{OD}$) reduces the momenta of all fragments, unbending of the DOD bond angle ($\theta_{\mathrm{DOD}}$) increases the angle between the two D$^+$ momenta ($\beta$), and alignment sees the angle between the \mbox{D-D} axis and the laser polarization axis ($\theta$) tend towards 0 and 180$^\circ$. \\
\begin{figure*}[t]
\centering
\includegraphics[width=19cm, trim={0cm 0cm 0cm 0cm}]{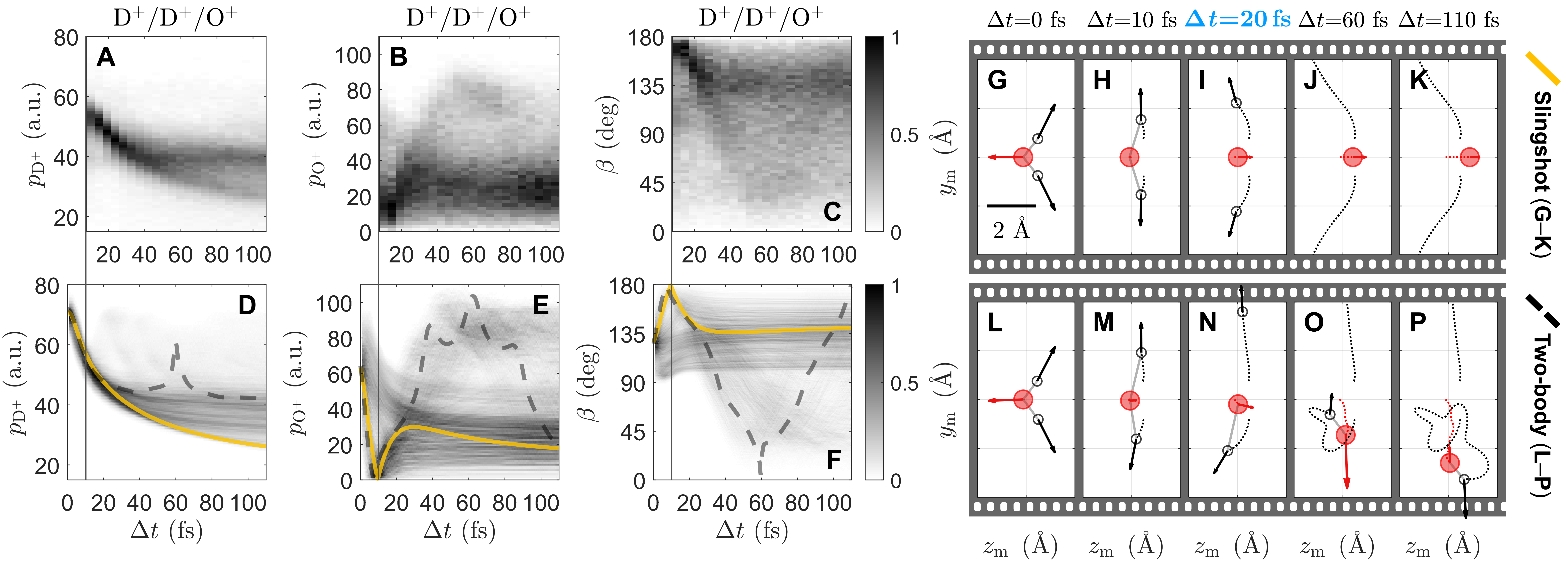}
\label{fig:FilmStrip}
\vspace{-0.2cm}
\caption{ \textbf{Filming nuclear motion in D$_2$O$^{2+}$.} (\textbf{A}-\textbf{C}) Measured distributions of deuteron momentum ($p_\mathrm{D^{+}}$), oxygen-ion momentum ($p_\mathrm{O^{+}}$), and momentum-frame bend-angle ($\beta$) plotted as a function of interpulse delay ($\Delta t$), detected in the triply charged three-body dissociation channel (D$^{+}$/D$^{+}$/O$^{+}$). 
(\textbf{D}-\textbf{F}) Theoretical distributions of $p_\mathrm{D^{+}}$, $p_\mathrm{O^{+}}$, and $\beta$ as a function of $\Delta t$.
(\textbf{G}-\textbf{K}) A series of film-strip plots for the class of trajectory labeled ``slingshot''. 
These plots displays both the position of the atoms (red solid circles for oxygen and black open circles for hydrogen) and their corresponding asymptotic momenta following formation of the trication. (\textbf{L}-\textbf{P}) Another series of film-strip plots for the class of trajectory labeled ``two-body''. 
The slingshot and two-body trajectories are highlighted in panels (D-F) as a solid yellow and dashed black line, respectively.}
\end{figure*}
\indent Stretching, unbending, and alignment are evidenced by the one-dimensional distributions shown in Figs.~2F, 2G, and 2H respectively.
Fig.~2F shows the total kinetic energy release (KER) decreasing from approximately 30 to 20~eV as pulse duration is increased from 6 to 19~fs.
This trend suggests that $r_\mathrm{OD}$ stretches as the pulse duration increases, reducing the KER of the Coulomb explosion, until a specific bond length is reached.
Fig.~2G shows $\beta$ rapidly increasing from approximately 140 to 180$^\circ$ as the pulse duration is increased from 6 to 10~fs, suggesting that the molecule may unbend in $<$10~fs; however, rapid unbending motion can cause $\beta$ to reach 180$^\circ$ even when $\theta_{\mathrm{DOD}}$ is less than 180$^\circ$.
Finally, Fig.~2H shows that $\theta$ grows more sharply peaked at 0 and 180$^\circ$ as pulse duration increases, suggesting dynamic alignment of the molecule's D-D axis with the laser polarization axis \cite{mccracken_ionization_2020,howard_strong-field_2021}. \\
\indent Next we compare the single-pulse momentum distributions to the double-pulse momentum distributions at an interpulse delay of 18 fs, the delay at which $R$ is maximized (Fig.~1C).
Figure~2E shows that ionization by pulse pairs leads to the same three intramolecular distortions as seen with single pulses.
\mbox{1-dimensional} distributions of KER, $\beta$, and $\theta$ for this particular interpulse delay are likewise reproduced in \mbox{Figs.~2F--2H}, also showing evidence for stretching, unbending, and alignment correlated with increased trication formation. \\
\\
\noindent \fontfamily{ptm}\selectfont \textbf{Recording and Modelling Intramolecular Motion} \\
\indent \fontfamily{ptm}\selectfont Ionizing with pulse pairs and tracking the CEI observables as a function of interpulse delay offers a wealth of information on the intramolecular dynamics leading to enhanced trication formation.
When paired with detailed \it{ab initio}\rm{} calculations, these observables serve as an unambiguous probe of state-selective interatomic motion.
In our theoretical treatment of the nuclear dynamics, the leading pulse in the pair promotes the Wigner phase space distribution of the ground vibrational state of neutral D$_2$O to the accurately computed potential energy surface of any one of the nine states of D$_2$O$^{2+}$ with two vacancies in the valence orbitals \cite{gervais_h2o2_2009,streeter_dissociation_2018,reedy_dissociation_2018,severt_step-by-step_2022}.
These nine states are detailed in Table~1 of Methods and Materials.
This phase space distribution is then propagated using classical trajectories on each of the nine potential energy surfaces for a time, $\Delta t$, before arrival of the second pulse and formation of the trication. 
Upon formation of the trication, the trajectories are continued under simple Coulomb repulsion of three singly-charged fragments, allowing the straightforward extraction of the asymptotic fragment momenta for each delay (see Methods and Materials for further detail on this procedure).
\\
\indent Comparing the experimental observables to their simulated counterparts allows us to identify which nuclear dynamics in D$_2$O$^{2+}$ ultimately contribute to the formation of D$_2$O$^{3+}$.
In \mbox{Figs.~3A--3C}, three experimental observables are plotted for all D$^+$/D$^+$/O$^+$ coincidences over interpulse delay: the magnitude of the deuteron momentum ($p_\mathrm{D^+}$) the magnitude of the oxygen-ion momentum ($p_\mathrm{O^+}$) and momentum-frame bend angle ($\beta$).
In \mbox{Figs.~3D--3F}, the same three observables are obtained from an ensemble of calculated trajectories and plotted similarly, reproducing many qualitative features of the data (see section 2.1 of the Supplementary Materials for details).\\
\indent In order to emphasize the distinct features within \mbox{Figs.~3D--3F}, two example trajectories are highlighted, one labeled ``slingshot'' and the other labeled ``two-body''.
These two trajectories are depicted in a series of film-strip plots in \mbox{Figs.~3G--3K} and \mbox{Figs.~3L--3P}, respectively.
These plots display the instantaneous molecular geometry and asymptotic fragment momenta for five particular values of $\Delta t$.
The first trajectory, labeled ``slingshot'', is an example of symmetric breakup into three bodies.
Here, both deuterons move symmetrically following double ionization and the molecule undergoes a rapid slingshot motion in which the bend-angle is inverted about the $z_\mathrm{m}$ axis.
As the molecule unbends and stretches, it briefly becomes linear at $\Delta t$ = 20~fs before re-bending the other way.
This kind of slingshot trajectory only occurs on three of the nine states of D$_2$O$^{2+}$, and it is most common in the relatively high-lying 2~$^1$A$_1$ state \cite{streeter_dissociation_2018,reedy_dissociation_2018} (see Table~1 for branching ratios).
The second trajectory, labeled ``two-body'', is an example of asymmetric breakup.
Here, only one deuteron is ejected following double ionization, leaving the other deuteron orbiting the oxygen atom as part of a rotationally and vibrationally hot OD$^+$ fragment.
This kind of two-body trajectory occurs on four of the nine states of D$_2$O$^{2+}$ and is most common in the lowest three states ($^3$B$_1$, 1~$^1$A$_1$, and $^1$B$_1$) \cite{streeter_dissociation_2018,reedy_dissociation_2018}.
\\
\indent Atomic motion is revealed in detail by comparing the data to the two trajectories highlighted in Fig.~3. 
For example, Fig.~3A displays $p_\mathrm{D^+}$ decreasing monotonically before bifurcating into two branches at later delays.
Each branch is well approximated by a highlighted trajectory in Fig.~3D, representing a deuteron that escaped via the slingshot trajectory or the deuteron that first escaped via the two-body trajectory.
In Fig.~3C, $\beta$ is centered at 180$^\circ$ after only 10 fs.
Figs.~3H and 3M demonstrate that this corresponds to a $\theta_{\mathrm{DOD}}$ of only 146$^\circ$ and 158$^\circ$, respectively.
In the slingshot trajectory, the molecule becomes linear at $\Delta t$~=~20 fs, pictured in Fig.~3I, at which time $\beta$ has already bent backwards, to 148$^\circ$.
The backwards bend also reverses the momentum of the oxygen ion, as seen in Figs.~3G--3K. 
Evidence for this reversal is seen in the earliest delays of Fig.~3B where the measured oxygen momentum appears to pass through zero.\\
\indent Additional two-body motion can also be revealed by comparing the data with the appropriate trajectory.
Fig.~3C displays a feature that bends rapidly to $\beta \sim$~0$^\circ$ and then unbends after about 60~fs.
The two-body trajectory in Fig.~3F reproduces this feature.
\mbox{Figs.~3L--3P} demonstrate how the bound deuteron rotates around the oxygen atom within 60~fs such that its momentum again aligns with the dissociating deuteron and yields $\beta$~=~0$^\circ$.
Likewise, in Fig.~2B, the faint cluster near $p_\mathrm{O^+}\sim$~80~a.u. at $\Delta t$~$\sim$~60~fs can also be attributed to this same motion in the two-body trajectory.
Here, both deuterons are oriented to act together in repulsing the oxygen ion, giving it maximal momentum (Fig.~3O).\\
\\
\noindent \fontfamily{ptm}\selectfont \textbf{Extracting the Critical Geometry} \\
\indent \fontfamily{ptm}\selectfont We can now determine which geometries are responsible for the enhancement in trication production observed at particular pulse-pairs delays.
The trication yield in Fig.~1C is maximal for $\Delta t$~=~18 fs.
Near this particular delay (at $\Delta t$~=~20~fs), the slingshot trajectory, for example, traverses through the geometry in which the molecule is briefly linear (Fig.~3I).
Here $r_\mathrm{OD}$ has stretched to approximately 2.2~$\text{\AA}$.
To fully explore each distorted molecular geometry and find its contribution to the enhancement, we examined all of the simulated trajectories that comprise Figs.~3D--3F. \\
\indent We extracted the critical geometry of EI from the trajectories by utilizing the fact that the observed enhancement is localized not only in time but in momentum and angle.
Using three observables ($\Delta t$, $\beta$, and $p_\mathrm{D^+}$) we constructed a histogram that localizes the ``enhancement volume'' in 3-dimensions.
An isointensity surface of this 3-dimensional volume at 50$\%$ maximum enhancement is depicted in Fig.~4A. 
To retrieve geometrical information at the moment of the enhancement, we propagated all trajectories through this 3-dimensional space and assigned each trajectory a weight per time-step based on the local value of the enhancement (see section 2.2 of the Supplementary Materials for details).
Two sample trajectories for each state are plotted traversing through this 3-dimensional space in Fig.~4A.
The three states that undergo rapid slingshot motion, bending backwards in $\leq$~20~fs, are highlighted in yellow, whereas all other states are depicted in black.
Fig.~4B shows the results of this analysis, plotting the combined contributions of all weighted trajectories in coordinates of $r_\mathrm{OD}$ and $\theta_{\mathrm{DOD}}$.
This reveals an enhancement for bond lengths between 1.8 and 2.5~$\text{\AA}$ as well as bend angles between 160 and 180$^\circ$, with the maximum of the distribution occurring at $r_\mathrm{OD}$~=~2.2~$\text{\AA}$ and $\theta_\mathrm{DOD}$~=~180$^\circ$. \\
\indent The inset of Fig.~4B reveals that it is only the states that undergo rapid slingshot motion (highlighted in yellow) that make a large contribution to this enhancement.
(Slower slingshot motion occurs more rarely: $<$1$\%$ of dissociations on the 1$^1$A$_1$ state. This state is responsible for the black trajectories that traverse through the enhancement volume, as seen in Fig.~4A and 4B.)
Accessing any one of the states in yellow ($^3$B$_1$, $^1$B$_1$, or 2~$^1$A$_1$) involves forming at least one vacancy in the 3a$_1$ molecular orbital while the 1b$_2$ orbital remains doubly occupied (see Fig.~1a for a schematic picture of these two orbitals).
This observation aligns with Walsh diagram rules and with our intuition: forming a vacancy in the 3a$_1$ orbital drives unbending motion in water, whereas a vacancy in the 1b$_2$ orbital drives rapid dissociation of the deuterons \cite{walsh_466_1953,streeter_dissociation_2018}.
Therefore, ionizing from the 3a$_1$ orbital while keeping the 1b$_2$ orbital intact allows substantial unbending prior to 3-body dissociation, permitting rapid slingshot motion.
The largest contribution to the enhancement is from the 2$^1$A$_1$ state, corresponding to a double vacancy in the 3a$_1$ molecular orbital.
Uniquely, 74$\%$ of all dissociations on this state undergo rapid slingshot motion, whereas this motion only occurs in 7$\%$ and 12$\%$ of dissociations on the other two states ($^3$B$_1$ and $^1$B$_1$ respectively). \\

\begin{figure}[H]
\centering
\includegraphics[width=7.2cm, trim={0cm 0cm 0cm 3.5cm}]{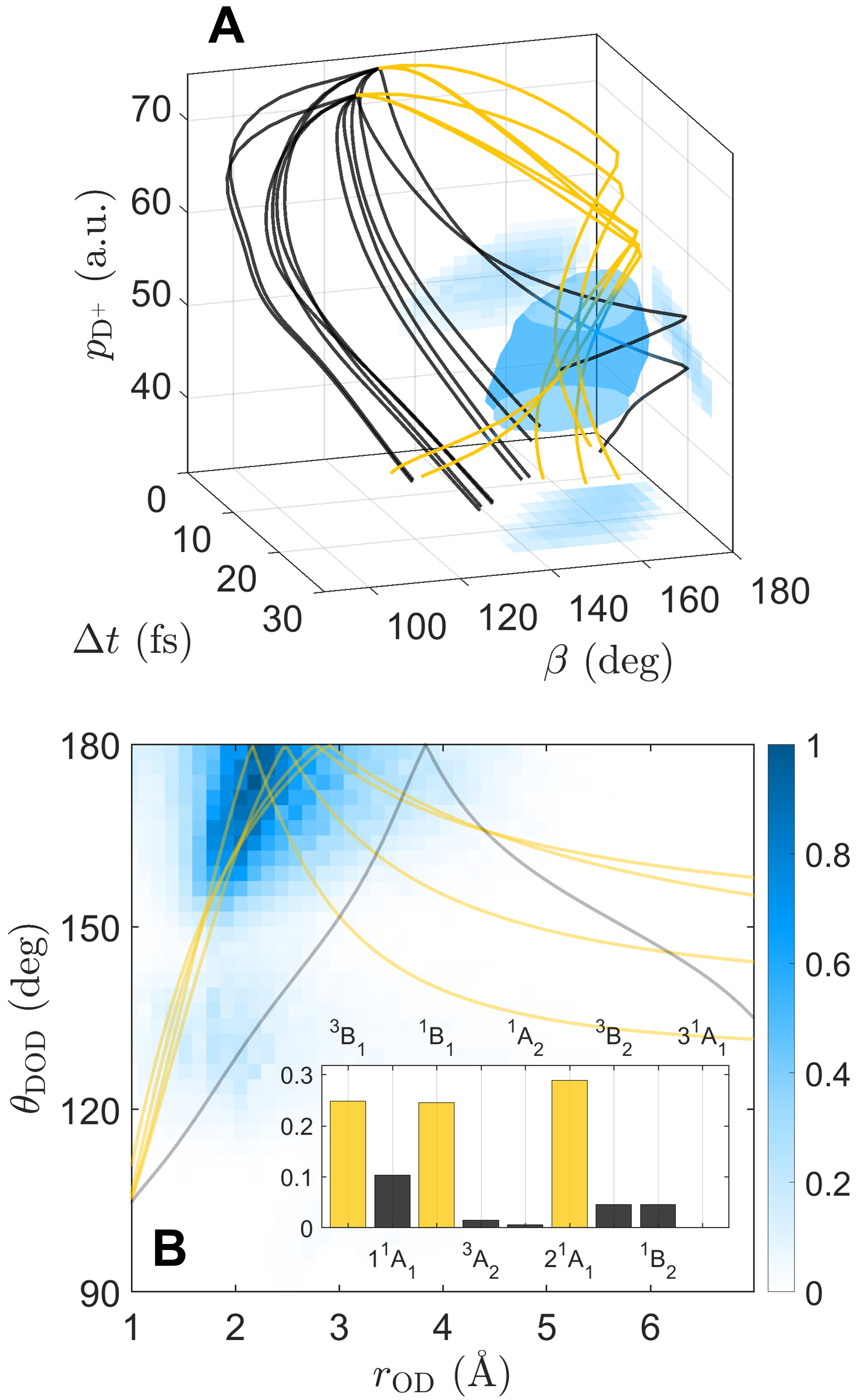}
\label{fig:Geometry}
\caption{ \textbf{Localizing the enhancement in D$_2$O$^{2+}$.} (\textbf{A}) A \mbox{3-dimensional} representation (in $p_\mathrm{D^+}$, $\beta$, and $\Delta t$) of the volume over which the enhanced trication production is within 50$\%$ of its maximum value (in semi-transparent cyan). 2-dimensional projections of the enhancement volume are displayed (in faded cyan) on the three walls of the plot. 
Two sample trajectories on each of the nine states of D$_2$O$^{2+}$ are also plotted within this 3-dimensional space. 
The trajectories on the states that undergo rapid slingshot motion are drawn in yellow, and those on the states that do not are drawn in black. 
(\textbf{B}) A 2-dimensional histogram of OD bond distance ($r_\mathrm{OD}$) and bend angle ($\theta_\mathrm{DOD}$) created by a weighted sum of the simulated trajectories.
Five sample trajectories from panel (A) are reproduced in panel (B). 
The weighted state populations are shown in the inset of panel (B). 
Those depicted in yellow undergo rapid slingshot motion and those in black do not.}
\end{figure}

\noindent \fontfamily{ptm}\selectfont \textbf{Modelling the EI Mechanism} \\
\indent \fontfamily{ptm}\selectfont We now attempt to model the enhancement mechanism that facilitates ionization of the third electron in D$_2$O$^{2+}$.
Models of EI in diatomic cations such as H$_2^{+}$ often invoke a 1-dimensional tunnelling picture in which the double-well potential is distorted by a static field \cite{xu_experimental_2015}.
In this picture, the presence of the downhill hydrogen suppresses the tunneling barrier for electrons localized on the uphill hydrogen.
The critical geometry is determined by balancing two competing factors: smaller bond lengths will cause greater barrier suppression but larger bond lengths will trap electronic population more effectively on the uphill hydrogen.
Here, we will invoke a similar tunneling picture to explain EI in D$_2$O$^{2+}$. \\
\indent A distinct feature of EI in H$_2^{+}$ is enhancement at large bond lengths ($>$5~$\text{\AA}$) due to a phenomenon known as Charge Resonance Enhanced Ionization (CREI)  \cite{trump_multiphoton_1999,trump_strong-field_1999,ergler_time-resolved_2005,ben-itzhak_elusive_2008,xu_experimental_2015}.
We have evidence that this charge resonance effect does not play a similar role in the enhancement in D$_2$O$^{2+}$.
Symmetric stretch in D$_2$O$^{2+}$ yields the asymptotic products D$^+$/D$^+$/O for all of the nine dicationic states considered here \cite{streeter_dissociation_2018}.
None of these states dissociate to O$^+$ as would be required for charge resonance effects to emerge \cite{mulliken_intensities_1939,zuo_charge-resonance-enhanced_1995}.\\
\indent The observed enhancement has a preferred polarization (Fig.~2H).
All trications, whether formed by pulse pairs or single pulses, have their D-D axis parallel to the laser polarization axis ($\hat{\epsilon}$).
For the pulse pairs, $\hat{\epsilon}$ refers to the polarization of the second pulse.
In both the single and double pulse data, the alignment preference appears as a distribution of D$^+$/D$^+$/O$^+$ coincidences peaked at $\theta$~=~0~and~180$^\circ$.
Utilizing the pulse pair data, the distribution in $\theta$ can be plotted as a function of delay, generating Fig.~5A.
Here, the alignment preference is localized only around the enhancement at $\Delta t$~=~18~fs. 
The distribution in $\theta$ becomes increasingly uniform far from this delay.
A tunneling picture of ionization suggests that, for a brief window of time around $\Delta t$~=~18~fs, the barrier to ionize is suppressed along $\theta$~=~0$^\circ$. \\
\indent To construct our model of EI, we generated the molecular electrostatic potential (MEP) during the enhancement using the critical geometry discussed in the previous section.
Modelling the MEP for the ground state of D$_2$O$^{3+}$ at the approximate critical geometry ($r_\mathrm{OD}=2.2~\text{\AA}$ and $\theta_\mathrm{DOD}=180^\circ$), subjecting it to our peak field strength (0.17~a.u.), and making two 1-dimensional cuts in $\theta$ ($\theta$~=~0$^\circ$ and $\theta$~=~90$^\circ$), yields Fig.~5B.
If we incorporate the ionization potential (IP) of D$_2$O$^{2+}$ ($\sim$40.7~eV) in Fig.~5, the $\theta$ dependence of the ionization barrier becomes apparent (see Methods and Materials for details on the calculation of IP).
When $\theta$~=~90$^\circ$, an electron localized near the oxygen atom must tunnel through a substantial barrier to ionize; however, when $\theta$~=~0$^\circ$, the barrier is nearly suppressed below the binding energy of the electron by the charge of the downhill deuteron, facilitating tunnelling if not over-the-barrier ionization.
Because the molecule is linear and stretched symmetrically, this same process may also occur on the opposite side of the molecule during the next half-cycle of the field.
The third ionization most likely creates a vacancy in the $\sigma$ orbital of the linear molecule (the 1b$_2$ orbital in C$_{2\mathrm{v}}$ symmetry), as this orbital has the largest value of electron density at the tunnelling barrier, and is doubly occupied in all three rapid slingshot states.
The additional degeneracy introduced by unbending and symmetric stretching \cite{streeter_dissociation_2018} may also supplement the tunneling current due to field-assisted couplings between states. \\
\indent In this model, the global minimum of the tunneling barrier occurs at $r_\mathrm{OD} \sim$~1.8~$\text{\AA}$, notably different than the critical geometry indicated by Fig.~4B.
This could indicate that the bond distances recovered in Fig.~4B are more likely the result of the particular trajectories launched by double ionization, rather than representative of the global optimum in bond length for EI.
This disparity does not occur in diatomic molecules because motion is only along one dimension, but it is a feature in polyatomics due to the increased degrees of freedom: constraints exist for EI in both bend angle and bond length.
\begin{figure*}
\centering
\includegraphics[width=15cm, trim={0cm 0cm 0cm 0cm}]{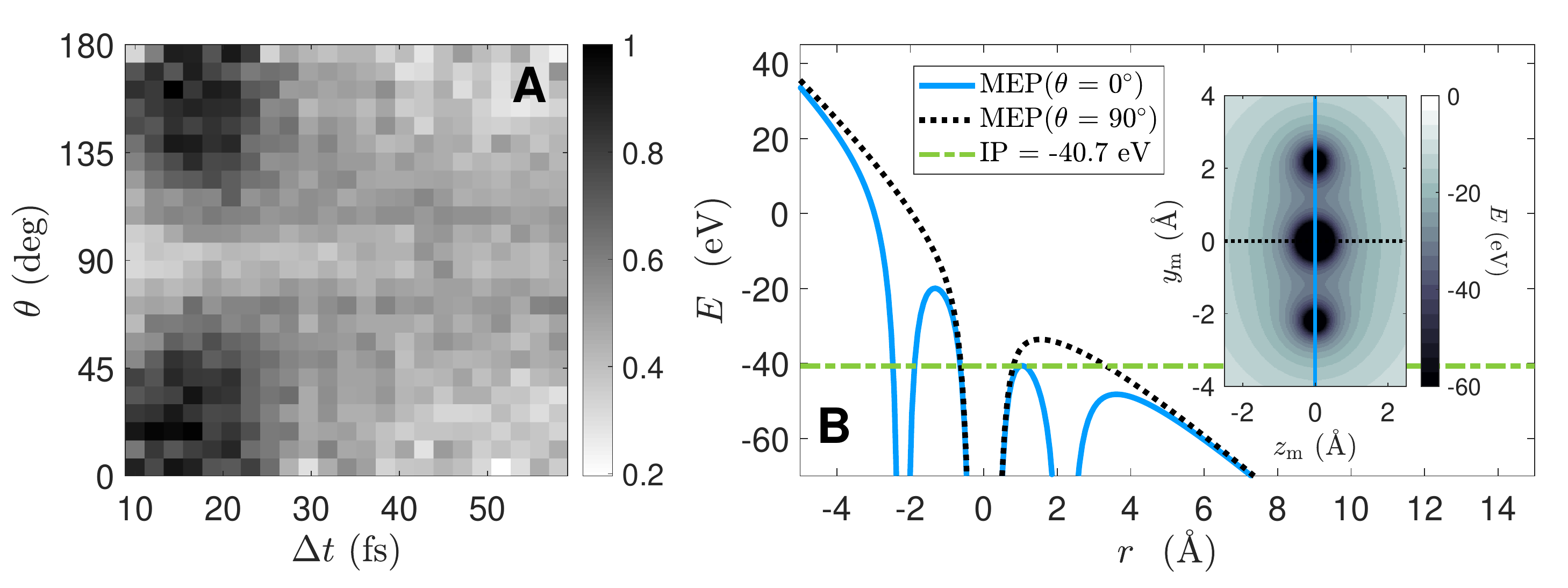}
\label{fig:Alignment}
\caption{ \textbf{A model of enhanced ionization in D$_2$O$^{2+}$.} (\textbf{A}) The distribution of alignment angle $\theta=\mathrm{arccos}(\hat{y}_{\mathrm{m}} \cdot \hat{\epsilon})$ for all D$^{+}$/D$^{+}$/O$^{+}$ coincidences as a function of interpulse delay ($\Delta t$). 
(\textbf{B}) The simulated molecular electrostatic potential (MEP) of D$_2$O$^{3+}$ for a linearly polarized DC field strength of 0.17 a.u. acting along two 1-dimensional cuts at $\theta=0^\circ$ (solid cyan line) and $\theta=90^\circ$ (dotted black line).
Also plotted is the simulated ionization potential (IP) of D$_2$O$^{2+}$ (dash-dotted green line).
The inset to panel (B) displays a top-down view of the 2-dimensional MEP (prior to distortion by the DC field) including lines to represent both 1-dimensional cuts: $\theta=0^\circ$ in solid cyan and $\theta=90^\circ$ in dotted black.}
\end{figure*}

\indent Fig.~2F shows that single-pulses produce significantly greater KER than double pulses at the optimal delay.
Single pulses with 19 or 40-fs duration produce a KER distribution peaked at 21.5~eV, while double pulses with an 18-fs delay have a KER distribution peaked at 17.5~eV.
If the single pulse KER is simply due to Coulomb repulsion in the linear molecule, it implies a symmetric bond length of only 1.66~$\text{\AA}$; however, this model excludes the kinetic energy acquired on the dication potential before the third ionization (see section 2.3 of the Supplementary Materials).  
In the two-pulse experiment where this effect was modeled, the molecules undergoing slingshot motion have $\sim$1.3~eV of kinetic energy just prior to the final ionization.
We have no comparable field-dressed prediction for the single-pulse experiment, but we can assume that it adds some energy (1-2~eV) to the Coulomb explosion.
Furthermore, the time-varying field may also distort the potentials due to phenomena such as bond softening \cite{bucksbaum_softening_1990,ben-itzhak_elusive_2008}: 
one of the field-dressed states of the water dication, for example, could drive motion that unbends the molecule with less rapid stretching.
Evidence for propagation on field-dressed states is already visible in the alignment preference for longer pulse durations (Fig.~2H) where the degree of alignment for pulses 19~fs and longer is far greater than for shorter pulses.
This suggests that dynamic alignment may have a substantial role in the EI process for long ($\tau \geq $19~fs) pulses \cite{mccracken_geometric_2017,mccracken_ionization_2020,howard_strong-field_2021}.\\
\\
\noindent \fontfamily{ptm}\selectfont \textbf{DISCUSSION} \\
\indent \fontfamily{ptm}\selectfont We have spatially and temporally resolved the rapid slingshot motion in D$_2$O that induces EI within 20~fs.
This motion is initiated by double ionization and creates, for a brief window in time, a linear DOD molecule with each OD bond distance stretched to about 2.2~$\text{\AA}$.
Within this window, a sufficiently strong and linearly polarized laser field oriented along the molecular axis can distort the molecular potential to facilitate ionization of a third electron and form D$_2$O$^{3+}$.
Electrons localized near the oxygen atom see a suppressed tunneling barrier due to the charge of the downhill deuteron, and are thus more easily ionized.
We see maximum EI when ionizing with 6-fs cross-polarized pulse pairs at pulse separations of 18~fs, but we also observe substantial EI when ionizing with single pulses of 19~fs duration or longer.
In both cases EI appears to be associated with unbending and symmetric stretching of the molecular bonds, but the KER are different, suggesting additional field-dressed dynamics in the case of single-pulse EI. \\
\indent Our measurements of the rapid intramolecular dynamics that produce EI in D$_2$O$^{2+}$ represent a new understanding of EI in polyatomics that can drastically affect the interpretation of CEI experiments.
The largest limitation of CEI is often ultrafast motion of light atomic species in the intermediate sequence of charge states that are formed en route to Coulomb explosion.
The distorting effects of EI are particularly egregious examples of this limitation.
Understanding this motion is therefore required before CEI can be applied to more complex molecular systems. \\
\indent This study shows not only how this motion can be measured in polyatomic molecules, but how the phenomenon of EI can be employed to do so.
We demonstrated how EI can act as a structural filter to highlight a single kind of interatomic motion in water.
It may be possible to tailor this EI process and select a subset of dynamics to highlight by changing the shape, intensity, or polarization of the ionizing laser field.
This capability may have already been seen in recent experiments examining an elusive roaming reaction in formaldehyde \cite{endo_capturing_2020}.
Control of EI in polyatomic molecules could therefore extend the applicability of laser-induced CEI, and may spark a renewed interest in its use to film ultrafast motion within polyatomic molecules. \\
\\
\noindent \fontfamily{ptm}\selectfont \textbf{MATERIALS AND METHODS} \\
\noindent \fontfamily{ptm}\selectfont \textbf{Producing Pulses of Variable Duration} \\
\indent \fontfamily{ptm}\selectfont In the first experimental scheme, single pulses of variable duration ($\tau$ = 6,~10,~19,~and~40~fs) and constant peak intensity ($I_0$~=~2$\times$10$^{15}$ W/cm$^2$) were used to triply ionize neutral D$_2$O.
Here, $\tau$ is a measure of the full width at half maximum (FWHM) of the temporal intensity profile for each pulse.
To create the 6-fs pulses, a 40-fs 800-nm 1-kHz Ti:sapphire laser pulse was spectrally broadened and chirped in a 1-m neon-filled hollow-core fiber (at 45 psi Ne) with a 250-$\mathrm{\mu}$m diameter.
The pulse (now with a central wavelength of 750-nm) was then recompressed using a series of bounces from two chirped mirror blocks.
The 10-fs pulses were created similarly but using a lower gas pressure (35 psi Ne) to elicit less spectral broadening.
The 19-fs pulses were created by positively chirping the 10-fs pulses using two fused silica wedges. 
Negatively chirping the 10-fs pulse to a duration of 19-fs yielded similar results.
The 40-fs pulse was supplied directly by the output of a Ti:sapphire laser oscillator and amplifier.
All peak intensities were kept constant by altering the energy per pulse using a series of neutral density pellicle filters with negligible dispersive effects.
For the 6 and 10-fs pulses, temporal characterization was performed via the dispersion scan method utilizing the aforementioned fused silica wedges to apply variable dispersion \cite{miranda_simultaneous_2012}.
For the longer pulse durations (19 and 40 fs), temporal characterization was performed via intensity autocorrelation \cite{diels_ultrashort_2006}. \\
\\
\noindent \fontfamily{ptm}\selectfont \textbf{Producing Pulse Pairs with Variable Delay} \\
\indent \fontfamily{ptm}\selectfont In the second experimental scheme, a pair of 6-fs 750-nm 1-kHz pulses with equal peak intensity ($I_0$~=~1$\times$10$^{15}$~W/cm$^2$) were used to triply ionize D$_2$O.
These pulses were generated by first creating a single 6-fs pulse (as described in the section above) before splitting this pulse using a Mach-Zehnder interferometer.
The interferometer splits the pulse into two pulses of equal intensity and variable delay. 
Each arm contained an additional polarizer at $\pm$45$^\circ$ to create a cross-polarized pair at the output directed along a common beam path.
Pulse characterization was performed via dispersion scan utilizing two BK7 wedges to apply variable amounts of dispersion \cite{miranda_simultaneous_2012}.
The interpulse delay was extracted with high precision from the spectral interference between the two beams in the unused output port of the interferometer. \\
\\
\noindent \fontfamily{ptm}\selectfont \textbf{Detection Geometry} \\
\indent \fontfamily{ptm}\selectfont In either of the two experimental schemes (described by the two preceding sections), the laser was ultimately directed into a vacuum chamber with a pressure of 6$\times$10$^{-10}$ Torr. 
The beam was then refocused back onto itself using a $f$=5~cm in-vacuum spherical metal mirror to form a focal spot of approximately 7~$\mu$m.
The chamber was then backfilled with a 50/50 mixture of gaseous H$_2$O and D$_2$O to a pressure of $\sim$1.5$\times$10$^{-9}$ Torr, such that $<$1 molecule was in the focus during each laser shot on average (operating at a repetition rate of 1 kHz).
As depicted schematically in the center of Fig.~1A, the laser induced multiple ionization in D$_2$O, causing rapid distortions to the molecular geometry that ultimately result in a Coulomb explosion into molecular fragments. 
These fragments were then accelerated toward a detector by a series of electrostatic plates held at high voltage. 
The detector was comprised of a triple-stack of microchannel plates and a Roentdek delay-line hex-anode \cite{jagutzki_multiple_2002}.
After post-processing of the electrical signals from the detector, this scheme yielded the full 3-dimensional momentum of each ionic fragment captured. \\
\\
\noindent \fontfamily{ptm}\selectfont \textbf{Recovering the Molecular Frame} \\
\indent \fontfamily{ptm}\selectfont When captured in coincidence, fragment momenta are initially in lab-frame coordinates but can be rotated into an experimentally recovered molecular frame by defining a new set of coordinates: $x_\mathrm{m}$, $y_\mathrm{m}$ and $z_\mathrm{m}$. 
Here, $\hat{z}_\mathrm{m}$ is defined as the bisector of the two D$^+$ momenta, $\hat{x}_\mathrm{m}$ as the cross product of the two D$^+$ momenta, and $\hat{y}_\mathrm{m}$ as the cross product between $\hat{z}_\mathrm{m}$ and $\hat{x}_\mathrm{m}$:

\begin{subequations}
\label{Eq_S1}
\begin{align}
    \hat{z}_\mathrm{m} & = 
    \left(\frac{ \vec{p}_\mathrm{D_{(1)}^{+}} }{|\vec{p}_\mathrm{D_{(1)}^{+}}|} + \frac{ \vec{p}_\mathrm{D_{(2)}^{+}} }{|\vec{p}_\mathrm{D_{(2)}^{+}}|}
    \right) /\hspace{1mm} 2\mathrm{cos}(\beta/2) \label{Eq_1a}\\
    \hat{x}_\mathrm{m} &= \left(\vec{p}_\mathrm{D_{(1)}^{+}} \times \vec{p}_\mathrm{D_{(2)}^{+}}\right) /\hspace{1mm} |\vec{p}_\mathrm{D_{(1)}^{+}}| \hspace{0.25mm} |\vec{p}_\mathrm{D_{(2)}^{+}}| \hspace{0.5mm} \mathrm{sin}(\beta) \label{Eq_1b}\\
    \hat{y}_\mathrm{m} &= \hat{z}_\mathrm{m} \times \hat{x}_\mathrm{m} \label{Eq_1c}\\
    \notag\\
    &\hspace{-4mm} \mathrm{where} \hspace{1mm} \beta = \mathrm{arccos}(\vec{p}_\mathrm{D_{(1)}^{+}} \cdot \vec{p}_\mathrm{D_{(2)}^{+}}) \notag
\end{align}
\end{subequations}

\noindent An important limitation of these coordinates is the lack of distinguishability between $+z_\mathrm{m}$ and $-z_\mathrm{m}$. 
Because the bisector of the two deuteron momenta is always defined as $+z_\mathrm{m}$, there is nothing distinguishing a molecule that has been inverted about $z_\mathrm{m}$ from one that has not. 
This inversion is only apparent after considering the evolution of certain time-resolved observables with pulse-pair separation. 
We see $\beta$ reach 180$^\circ$ before returning to more acute angles, indicating backward bending.
We also see the magnitude of the oxygen-ion momentum ($p_\mathrm{O^+}$) pass through 0, indicating a change of sign. (See section 1.1 of the Supplementary Materials for further detail on the limitations of Eqs.~\ref{Eq_1a}--\ref{Eq_1c}.)\\

\begin{table*}[t]
\centering
\begin{tabular}{ |c|c|c|c|c|c| } 
 \hline
 \textbf{$\bm{\Delta E}$ (eV)} & \textbf{C$\bm{_\mathrm{2v}}$ Symmetry} & \textbf{C$_\mathrm{s}$ Symmetry} & \textbf{Orbital Configuration} & \textbf{2/3-Body Branching Ratio ($\bm{\%}$)} & \textbf{Rapid Slingshot} \\ 
 \hline
 40.3 & $^3$B$_1$ & $^3$A$^{\prime\prime}$ & (3a$_1$)$^{-1}$(1b$_1$)$^{-1}$ & 93.0 / 7.0 & Yes \\ 
 \hline
 41.4 & 1 $^1$A$_1$ & 1 $^1$A$^{\prime}$ & (1b$_1$)$^{-2}$ & 99.4 / 0.6 & No\\ 
 \hline
 42.8 & $^1$B$_1$ & 1 $^1$A$^{\prime\prime}$  & (3a$_1$)$^{-1}$(1b$_1$)$^{-1}$ & 87.7 / 12.3 & Yes \\
 \hline
 44.3 & $^3$A$_2$ & 2 $^3$A$^{\prime\prime}$ & (1b$_2$)$^{-1}$(1b$_1$)$^{-1}$ & 0.0 / 100.0 & No \\
 \hline
 46.0 & $^1$A$_2$ & 2 $^1$A$^{\prime\prime}$ & (1b$_2$)$^{-1}$(1b$_1$)$^{-1}$ & 0.0 / 100.0 & No \\
 \hline
 46.0 & 2 $^1$A$_1$ & 2 $^1$A$^{\prime}$ & (3a$_1$)$^{-2}$ & 26.3 / 73.7 & Yes \\
 \hline
 46.3 & $^3$B$_2$ & 1 $^3$A$^{\prime}$ & (1b$_2$)$^{-1}$(3a$_1$)$^{-1}$ & 0.0 / 100.0 & No \\
 \hline
 48.4 & $^1$B$_2$ & 3 $^1$A$^{\prime}$ & (1b$_1$)$^{-1}$(3a$_1$)$^{-1}$ & 0.0 / 100.0 & No \\
 \hline
 50.3 & 3 $^1$A$_1$ & 3 $^1$A$^{\prime\prime}$ & (1b$_2$)$^{-2}$ & 0.0 / 100.0 & No \\
 \hline
\end{tabular}
\caption{ \textbf{Electronic states of D$_2$O$^{2+}$.} \label{Tab_1} A tabulated list of all nine states of D$_2$O$^{2+}$ that correspond to the removal of two electrons from any combination of the valence orbitals: (1b$_2$)$^2$(3a$_1$)$^2$(1b$_1$)$^2$ as labeled by C$_{\mathrm{2v}}$ symmetry. For each state, the energy, symmetry, orbital configuration, 2/3-body branching ratio, and existence of rapid slingshot trajectories are listed. Here, the energy is written in terms of $\Delta E$, the energy difference between the neutral ground state of D$_2$O at the equilibrium geometry and the Franck-Condon point of a given state. These energies come directly from the potential energy surfaces calculated by Gervais {\it et al.} \cite{gervais_h2o2_2009} and Streeter {\it et al.}  \cite{streeter_dissociation_2018}. The 2/3-body branching ratios were found via simulation after propagating 2048 classical trajectories on each surface. ``Rapid Slingshot'' refers to the whether or not each state permits a 3-body dissociation trajectory that inverts the $z_\mathrm{m}$ axis of the molecule within 20~fs.}
\end{table*}

\noindent \fontfamily{ptm}\selectfont \textbf{Simulating the Dynamics of D$_2$O$^{2+}$ and D$_2$O$^{3+}$} \\
\indent \fontfamily{ptm}\selectfont To model the interatomic dynamics of D$_2$O following double ionization, we simulated the motion of a nuclear wavepacket propagating on the dication potential energy surfaces semi-classically  in the same way as was done in references  \cite{gervais_h2o2_2009} and \cite{streeter_dissociation_2018}, by using classical trajectories whose initial conditions are given by the Wigner phase space distribution of the initial vibrational state.
The Wigner distribution in the harmonic approximation used here is
\begin{equation}
W(\mathbf{Q},\mathbf{P}) = \frac{1}{ \left(\pi \hbar  \right)^{3} } \prod_{j=1}^{3}   \exp \left[{-\frac{\omega_j}{\hbar} Q_j^2 -\frac{P_j^2}{\hbar \omega}}  \right] \, ,
\label{eq:Wigner}
\end{equation}
where $Q_j$ and $P_j$ are the coordinates and momenta respectively of the three normal modes, and $\omega_j$ are the associated frequencies.  The normal modes were calculated in a complete active space self-consistent field (CASSCF) calculation with the same active space used in the calculations of the dication potential surfaces described below.  This semiclassical phase space distribution was propagated by sampling it to initiate a set of 2048 classical trajectories on each of the nine states of D$_2$O$^{2+}$.
The vertical transition energies, symmetries, orbital vacancies in the dominant configuration at the equilibrium geometry, and branching ratios for all of these states are displayed in Table~1. \\
\indent Eight of the three-dimensional potential energy surfaces used for this simulation were calculated by Gervais {\it et al.}  \cite{gervais_h2o2_2009} and the ninth, the $3^1A_1$ state, was calculated by Streeter {\it et al.}  \cite{streeter_dissociation_2018}.  
Briefly, the potentials were produced by internally contracted multi-reference configuration interaction (icMRCI) calculations at the configuration interaction singles and doubles (CISD) level including the Davidson correction for quadruple excitations.  The calculations, which were performed on extensive grids of geometries, employed the cc-pVTZ Dunning correlation consistent basis~\cite{dunning_gaussian_1989} and were based on a CASSCF reference space.   
For example in the case of the $3^1A_1$ state,  the active space used orbitals from 
calculations on the lowest $^3\textrm{B}_1$ state in C$_\textrm{s}$ symmetry with one a$^\prime$ orbital frozen and
six electrons in five a$^\prime$ and two a$^{\prime \prime}$ orbitals.  These accurate \textit{ab initio} surfaces calculated with MOLPRO~\cite{MOLPRO-WIREs,MOLPRO_brief} were fit to a linear combination of
100 basis functions that represent the Coulomb and
polarization interactions at intermediate and long interatomic
distances together with screened Coulomb and multipole interactions at
short distances as described in reference \cite{gervais_h2o2_2009}. \\
\indent The reliability of these potential surfaces and the Wigner phase space propagation on them for dissociative dynamics has been established in detailed comparisons with experimental momentum imaging experiments on one-photon double photoionization of H$_2$O and D$_2$O \cite{gervais_h2o2_2009,streeter_dissociation_2018,reedy_dissociation_2018,severt_step-by-step_2022}.  This treatment of the nuclear dynamics closely reproduces experimental final momentum distributions for the three-body breakup channels as well as internal energy distributions in the two-body channels in those studies.  Those benchmark comparisons underlie our confidence in the present investigation in using these detailed dynamics to interpret the experimental trajectories as described in the main article and below.\\
\indent After propagating on a given state of D$_2$O$^{2+}$ for a time commensurate with the interpulse delay, $\Delta t$, the formation of the D$_2$O$^{3+}$ and the resulting Coulomb explosion were modelled by an instantaneous transition (preserving the positions and momenta of the particles on the dication states) to a purely repulsive potential, $V$.
This potential simply represents mutual Coulomb repulsion between three single charges:
\begin{equation}
    V = \frac{q_1 q_2}{|\vec{r}_1-\vec{r}_2|} + \frac{q_1 q_3}{|\vec{r}_1-\vec{r}_3|} + \frac{q_2 q_3}{|\vec{r}_2-\vec{r}_3|}
\end{equation}
where $q_i$ is the net charge on each fragment ($q_i$~=~1 in atomic units), $\vec{r}_i$ is the three-dimensional position vector of each fragment, and $i=1,2,3$ correspond to D$^+_{(1)}$, D$^+_{(2)}$ and O$^+$ respectively.
The inherent timing ambiguity due to the finite width of the two pulses in the pair ($\tau$~=~6~fs) was accounted for by ``blurring'' the dication dynamics by $\pm$3~fs: that is, randomly shifting the timing of all 2048 trajectories forward and back by an amount between 0 and 3~fs and averaging over the results. \\
\\
\noindent \fontfamily{ptm}\selectfont \textbf{Tunneling Simulations} \\
\indent \fontfamily{ptm}\selectfont The molecular electrostatic potential (MEP) for the ground quartet state of D$_2$O$^{3+}$ displayed in Fig.~5B (inset) was generated neglecting exchange interactions using restricted open-shell Hartree-Fock (ROHF) theory in GAMESS \cite{GAMESS} with a 6-31G Gaussian basis set. 
A DC electric field of strength $\varepsilon_0$~=~0.17~a.u. (I$_0$~=~1$\times$10$^{15}$ W/cm$^2$) was then applied to this MEP along the $y_\mathrm{m}$ and $z_\mathrm{m}$ axes in order to yield the tunneling pictures in Fig.~5B for $\theta$~=~0$^\circ$ and 90$^\circ$ respectively. \\
\indent The ionization potential (IP) plotted in Fig.~5B as a horizontal line is from a CASSCF calculation of the difference between the energy of the ground state ($^3\Sigma^-$) of H$_2$O$^{2+}$ and that of H$_2$O$^{3+}$ ($^4\Sigma^-$) under the influence of an applied field of $\varepsilon_0$~=~0.17~a.u.  
In these calculations, performed with the Psi4 \cite{Psi4_2020} suite of codes, the Gaussian basis was again the cc-pVTZ Dunning correlation consistent basis~\cite{dunning_gaussian_1989}, and the OD distance was either 2 or 2.5~$\text{\AA}$.
Results were interpolated to yield the IP at 2.2~$\text{\AA}$.
The two tightest a$_1$ orbitals were doubly occupied in these calculations and the active space for calculations on both systems consisted of 2 a$_1$ 2 b$_1$ and 2 b$_2$ orbitals.
Correlated calculations like this give ionization energies in the presence of an E-field several eV larger than ROHF descriptions.
The difference evidently arises because CASSCF calculations can allow orbitals in correlating configurations to occupy the vicinity of the down-field deuteron, while single configuration-calculations spread the highest occupied orbital over both the oxygen and deuteron.
As noted in the main text, these estimates of the ionization energy, and therefore the energy in a one-dimensional model at which tunneling or over-barrier ionization takes place, are very approximate.
The calculations are complicated in particular by the difference in the electron correlation energies of the two systems, because the trication is dramatically less correlated than the dication.
Additionally, electronic states in intense fields have finite lifetimes and therefore energy widths, which are neglected in these calculations.
As mentioned in the main text, even if these computational estimates of the ionization energy and barrier height are off by one or two eV, the proposed mechanism for EI remains reasonable. \\

\nocite{miranda_simultaneous_2012,diels_ultrashort_2006,jagutzki_multiple_2002,dunning_gaussian_1989,MOLPRO-WIREs,MOLPRO_brief,GAMESS,Psi4_2020}

\noindent \fontfamily{ptm}\selectfont \textbf{REFERENCES}
\fontfamily{ptm}\selectfont

\bibliographystyle{Science}

\begin{thebibliography}{10}

\bibitem{brixner_two-dimensional_2005}
T.~Brixner, {\it et~al.\/}, {\it Nature\/} {\bf 434}, 625 (2005).

\bibitem{schwalb_base_2008}
N.~K. Schwalb, F.~Temps, {\it Science\/} {\bf 322}, 243 (2008).

\bibitem{prokhorenko_coherent_2006}
V.~I. Prokhorenko, {\it et~al.\/}, {\it Science\/} {\bf 313}, 1257 (2006).

\bibitem{dwyer_femtosecond_2006}
J.~R. Dwyer, {\it et~al.\/}, {\it Phil. Trans. R. Soc. A.\/} {\bf 364}, 741
  (2006).

\bibitem{barty_molecular_2013}
A.~Barty, J.~Küpper, H.~N. Chapman, {\it Annu. Rev. Phys. Chem.\/} {\bf 64},
  415 (2013).

\bibitem{ivanov_concluding_2021}
M.~Ivanov, {\it Faraday Discuss.\/} {\bf 228}, 622 (2021).

\bibitem{yang_imaging_2018}
J.~Yang, {\it et~al.\/}, {\it Science\/} {\bf 361}, 64 (2018).

\bibitem{wolf_photochemical_2019}
T.~J.~A. Wolf, {\it et~al.\/}, {\it Nat. Chem.\/} {\bf 11}, 504 (2019).

\bibitem{yang_simultaneous_2020}
J.~Yang, {\it et~al.\/}, {\it Science\/} {\bf 368}, 885 (2020).

\bibitem{meckel_laser-induced_2008}
M.~Meckel, {\it et~al.\/}, {\it Science\/} {\bf 320}, 1478 (2008).

\bibitem{wolter_ultrafast_2016}
B.~Wolter, {\it et~al.\/}, {\it Science\/} {\bf 354}, 308 (2016).

\bibitem{minitti_imaging_2015}
M.~Minitti, {\it et~al.\/}, {\it Phys. Rev. Lett.\/} {\bf 114}, 255501 (2015).

\bibitem{kim_direct_2015}
K.~H. Kim, {\it et~al.\/}, {\it Nature\/} {\bf 518}, 385 (2015).

\bibitem{glownia_self-referenced_2016}
J.~Glownia, {\it et~al.\/}, {\it Phys. Rev. Lett.\/} {\bf 117}, 153003 (2016).

\bibitem{li_time-resolved_2008}
W.~Li, {\it et~al.\/}, {\it Science\/} {\bf 322}, 1207 (2008).

\bibitem{worner_conical_2011}
H.~J. Wörner, {\it et~al.\/}, {\it Science\/} {\bf 334}, 208 (2011).

\bibitem{he_monitoring_2018}
L.~He, {\it et~al.\/}, {\it Nat. Commun.\/} {\bf 9}, 1108 (2018).

\bibitem{pitzer_direct_2013}
M.~Pitzer, {\it et~al.\/}, {\it Science\/} {\bf 341}, 1096 (2013).

\bibitem{kunitski_observation_2015}
M.~Kunitski, {\it et~al.\/}, {\it Science\/} {\bf 348}, 551 (2015).

\bibitem{fehre_enantioselective_2019}
K.~Fehre, {\it et~al.\/}, {\it Sci. Adv.\/} {\bf 5}, eaau7923 (2019).

\bibitem{endo_capturing_2020}
T.~Endo, {\it et~al.\/}, {\it Science\/} {\bf 370}, 1072 (2020).

\bibitem{erk_imaging_2014}
B.~Erk, {\it et~al.\/}, {\it Science\/} {\bf 345}, 288 (2014).

\bibitem{liekhus-schmaltz_ultrafast_2015}
C.~E. Liekhus-Schmaltz, {\it et~al.\/}, {\it Nat. Commun.\/} {\bf 6}, 8199
  (2015).

\bibitem{zeller_imaging_2016}
S.~Zeller, {\it et~al.\/}, {\it Proc. Natl. Acad. Sci. U.S.A.\/} {\bf 113},
  14651 (2016).

\bibitem{rudenko_femtosecond_2017}
A.~Rudenko, {\it et~al.\/}, {\it Nature\/} {\bf 546}, 129 (2017).

\bibitem{boll_x-ray_2022}
R.~Boll, {\it et~al.\/}, {\it Nat. Phys.\/} {\bf 18}, 423 (2022).

\bibitem{herwig_imaging_2013}
P.~Herwig, {\it et~al.\/}, {\it Science\/} {\bf 342}, 1084 (2013).

\bibitem{reedy_dissociation_2018}
D.~Reedy, {\it et~al.\/}, {\it Phys. Rev. A\/} {\bf 98}, 053430 (2018).

\bibitem{severt_step-by-step_2022}
T.~Severt, {\it et~al.\/}, {\it Nat. Commun.\/} {\bf 13}, 5146 (2022).

\bibitem{yamanouchi_visualizing_2012}
A.~Matsuda, M.~Fushitani, E.~J. Takahashi, A.~Hishikawa, {\it Multiphoton
  {Processes} and {Attosecond} {Physics}\/}, K.~Yamanouchi, M.~Katsumi, eds.
  (Springer Berlin Heidelberg, Berlin, Heidelberg, 2012), vol. 125, pp.
  317--322.

\bibitem{vager_coulomb_1989}
Z.~Vager, R.~Naaman, E.~P. Kanter, {\it Science\/} {\bf 244}, 426 (1989).

\bibitem{stapelfeldt_wave_1995}
H.~Stapelfeldt, E.~Constant, P.~B. Corkum, {\it Phys. Rev. Lett.\/} {\bf 74},
  3780 (1995).

\bibitem{fukuzawa_real-time_2019}
H.~Fukuzawa, {\it et~al.\/}, {\it Nat. Commun.\/} {\bf 10}, 2186 (2019).

\bibitem{rudenko_real-time_2006}
A.~Rudenko, {\it et~al.\/}, {\it Chem. Phys.\/} {\bf 329}, 193 (2006).

\bibitem{bocharova_time-resolved_2011}
I.~A. Bocharova, {\it et~al.\/}, {\it Phys. Rev. A\/} {\bf 83}, 013417 (2011).

\bibitem{legare_laser_2005}
F.~Légaré, {\it et~al.\/}, {\it Phys. Rev. A\/} {\bf 71}, 013415 (2005).

\bibitem{legare_laser_2006}
F.~Légaré, K.~F. Lee, A.~D. Bandrauk, D.~M. Villeneuve, P.~B. Corkum, {\it J.
  Phys. B\/} {\bf 39}, S503 (2006).

\bibitem{seideman_role_1995}
T.~Seideman, M.~Y. Ivanov, P.~B. Corkum, {\it Phys. Rev. Lett.\/} {\bf 75},
  2819 (1995).

\bibitem{zuo_charge-resonance-enhanced_1995}
T.~Zuo, A.~D. Bandrauk, {\it Phys. Rev. A\/} {\bf 52}, R2511 (1995).

\bibitem{bocharova_charge_2011}
I.~Bocharova, {\it et~al.\/}, {\it Phys. Rev. Lett.\/} {\bf 107}, 063201
  (2011).

\bibitem{liu_charge_2015}
H.~Liu, {\it et~al.\/}, {\it Chinese Physics Letters\/} {\bf 32}, 063301
  (2015).

\bibitem{wu_probing_2012}
J.~Wu, {\it et~al.\/}, {\it Nat. Commun.\/} {\bf 3}, 1113 (2012).

\bibitem{ibrahim_h_2018}
H.~Ibrahim, C.~Lefebvre, A.~D. Bandrauk, A.~Staudte, F.~Légaré, {\it J. Phys.
  B\/} {\bf 51}, 042002 (2018).

\bibitem{bucksbaum_softening_1990}
P.~H. Bucksbaum, A.~Zavriyev, H.~G. Muller, D.~W. Schumacher, {\it Phys. Rev.
  Lett.\/} {\bf 64}, 1883 (1990).

\bibitem{zhao_strong-field-induced_2019}
S.~Zhao, {\it et~al.\/}, {\it Phys. Rev. A\/} {\bf 99}, 053412 (2019).

\bibitem{mccracken_geometric_2017}
G.~A. McCracken, A.~Kaldun, C.~Liekhus-Schmaltz, P.~H. Bucksbaum, {\it J. Chem.
  Phys.\/} {\bf 147}, 124308 (2017).

\bibitem{mccracken_ionization_2020}
G.~A. McCracken, P.~H. Bucksbaum, {\it J. Chem. Phys.\/} {\bf 152}, 134308
  (2020).

\bibitem{cheng_momentum-resolved_2020}
C.~Cheng, {\it et~al.\/}, {\it Phys. Rev. A\/} {\bf 102}, 052813 (2020).

\bibitem{howard_strong-field_2021}
A.~J. Howard, {\it et~al.\/}, {\it Phys. Rev. A\/} {\bf 103}, 043120 (2021).

\bibitem{cheng_strong-field_2021}
C.~Cheng, {\it et~al.\/}, {\it Phys. Rev. A\/} {\bf 104}, 023108 (2021).

\bibitem{allum_multi-particle_2021}
F.~Allum, {\it et~al.\/}, {\it J. Phys. Chem. Lett.\/} {\bf 12}, 8302 (2021).

\bibitem{trump_multiphoton_1999}
C.~Trump, H.~Rottke, W.~Sandner, {\it Phys. Rev. A\/} {\bf 59}, 2858 (1999).

\bibitem{trump_strong-field_1999}
C.~Trump, H.~Rottke, W.~Sandner, {\it Phys. Rev. A\/} {\bf 60}, 3924 (1999).

\bibitem{ergler_time-resolved_2005}
T.~Ergler, {\it et~al.\/}, {\it Phys. Rev. Lett.\/} {\bf 95}, 093001 (2005).

\bibitem{ben-itzhak_elusive_2008}
I.~Ben-Itzhak, {\it et~al.\/}, {\it Phys. Rev. A\/} {\bf 78}, 063419 (2008).

\bibitem{xu_experimental_2015}
H.~Xu, F.~He, D.~Kielpinski, R.~Sang, I.~Litvinyuk, {\it Scientific Reports\/}
  {\bf 5}, 13527 (2015).

\bibitem{legare_time-resolved_2003}
F.~L\'egar\'e, {\it et~al.\/}, {\it Phys. Rev. Lett.\/} {\bf 91}, 093002
  (2003).

\bibitem{matsuda_time-resolved_2014}
A.~Matsuda, E.~J. Takahashi, A.~Hishikawa, {\it Journal of Electron
  Spectroscopy and Related Phenomena\/} {\bf 195}, 327 (2014).

\bibitem{koh_ionization_2020}
S.~Koh, K.~Yamazaki, M.~Kanno, H.~Kono, K.~Yamanouchi, {\it Chem. Phys.
  Lett.\/} {\bf 742}, 137165 (2020).

\bibitem{jagutzki_multiple_2002}
O.~Jagutzki, {\it et~al.\/}, {\it IEEE Transactions on Nuclear Science\/} {\bf
  49}, 2477 (2002).

\bibitem{gervais_h2o2_2009}
B.~Gervais, {\it et~al.\/}, {\it J. Chem. Phys.\/} {\bf 131}, 024302 (2009).

\bibitem{streeter_dissociation_2018}
Z.~L. Streeter, {\it et~al.\/}, {\it Phys. Rev. A\/} {\bf 98}, 053429 (2018).

\bibitem{walsh_466_1953}
A.~D. Walsh, {\it Journal of the Chemical Society (Resumed)\/} p. 2260 (1953).

\bibitem{mulliken_intensities_1939}
R.~S. Mulliken, {\it J. Chem. Phys.\/} {\bf 7}, 20 (1939).

\bibitem{miranda_simultaneous_2012}
M.~Miranda, T.~Fordell, C.~Arnold, A.~L’Huillier, H.~Crespo, {\it Optics
  Express\/} {\bf 20}, 688 (2012).

\bibitem{diels_ultrashort_2006}
J.-C. Diels, W.~Rudolph, {\it Ultrashort laser pulse phenomena\/}, Optics and
  photonics (Elsevier / Academic Press, Amsterdam; Boston, 2006), second edn.

\bibitem{dunning_gaussian_1989}
T.~H. Dunning, {\it J. Chem. Phys.\/} {\bf 90}, 1007 (1989).

\bibitem{MOLPRO-WIREs}
H.-J. Werner, P.~J. Knowles, G.~Knizia, F.~R. Manby, M.~Sch{\"u}tz, {\it WIREs
  Comput Mol Sci\/} {\bf 2}, 242 (2012).

\bibitem{MOLPRO_brief}
H.-J. Werner, {\it et~al.\/}, Molpro, version 2015.1, a package of ab initio
  programs (2015). See http://www.molpro.net.

\bibitem{GAMESS}
G.~M.~J. Barca, {\it et~al.\/}, {\it J. Chem. Phys.\/} {\bf 152}, 154102
  (2020).

\bibitem{Psi4_2020}
D.~G.~A. Smith, {\it et~al.\/}, {\it J. Chem. Phys.\/} {\bf 152}, 184108
  (2020).

\end{thebibliography}

\vspace{0.2cm} \noindent \small{\textbf{Acknowledgements:}}
The authors thank M. Spanner for helpful discussion.
\textbf{Funding:} A.J.H., M.B., R.F., F.A., J.L.R., G.A.M., I.G., and P.H.B. were supported by the National Science Foundation.
A.J.H. was additionally supported under a Stanford Graduate Fellowship as the 2019 Albion Walter Hewlett Fellow.
R.F. acknowledges support from the Linac Coherent Light Source, SLAC National Accelerator Laboratory, which is supported by the US Department of Energy, Office of Science, Office of Basic Energy Sciences, under contract no. DE-AC02-76SF00515.
I.G. was additionally supported by an NDSEG Fellowship.
C.C. and T.W. were supported by the US Department of Energy under Award No. DE-FG02-08ER15984. 
Work at Lawrence Berkeley National Laboratory (LBNL) was performed under the auspices of the U.S. Department of Energy (DOE), Office of Science, Office of Basic Energy Sciences, Chemical Sciences, Geosciences, and Biosciences Division under Contract No. DE-AC02-05CH11231, using the National Energy Research Computing Center (NERSC), a DOE Office of Science User Facility, and the Lawrencium computational cluster resource provided by LBNL.
\textbf{Author contributions:} A.J.H., M.B., and P.H.B. conceptualized the experiment.
A.J.H., M.B., J.L.R., G.A.M., and I.G. conducted the experimental investigation.
A.J.H., Z.L.S., and C.W.M. performed the theoretical investigation.
A.J.H. performed formal analysis of the experimental data.
A.J.H., M.B., R.F., G.A.M., and P.H.B. designed the experimental methodology.
A.J.H., Z.L.S., R.R.L., and C.W.M. designed the theoretical methodology.
A.J.H. and P.H.B. wrote the original draft of the paper.
A.J.H., M.B., C.C., R.F., F.A., J.L.R., C.W.M., T.W., and P.H.B. reviewed and edited the paper.
\textbf{Competing interests:} The authors declare no competing interests.
\textbf{Data and materials availability:} All data is available in the manuscript or the supplementary materials.
\\
\\
\noindent \fontfamily{ptm}\selectfont \textbf{SUPPLEMENTARY MATERIALS} \\
\fontfamily{ptm}\selectfont Supplementary Text \\
Fig.~S1. Comparing theoretical D$_2$O$^{2+}$ dynamics to the data. \\
Fig.~S2. Delay-dependent yield of D$_2$O$^{3+}$. \\
Fig.~S3. KER of D$_2$O$^{3+}$ with and without intermediate dynamics. \\
Table~S1. Theoretical populations of D$_2$O$^{2+}$ extracted from the data. \\

\end{multicols}


\end{document}


\maketitle

\vspace{-1.25cm}
\tableofcontents

\vspace{0.75cm}
\section{Experimental Methods}
\subsection{The Jacobian Factor}
\indent Equations 1a--1c break down when $\beta$~=~180$^\circ$. 
Here there is nothing to distinguish the $x_\mathrm{m}$ axis from the $z_\mathrm{m}$ axis since the momenta of the deuterons are exactly back-to-back. 
This is not a problem experimentally, however, because the probability of detecting two deuterons with exactly opposite momentum vectors is vanishingly small. 
This fact can be most easily visualized by imagining a 3-dimensional space (in $x$,~$y$,~and~$z$) where the normalized momentum vector of one deuteron always defines the -$z$ axis.
The normalized momentum vector of the other deuteron could therefore exist anywhere on the surface of a 3-dimensional unit sphere.
Finding the total distribution in $\beta$ requires integrating over the surface area of this sphere.
To do so, it is useful to use spherical coordinates where the differential surface area of the unit sphere (d$A$) can be written as d$A$~=~2$\pi~\mathrm{cos}(\phi)$~d$\phi$, where $\phi$ is the polar angle to the $z$-axis.
Here, $\mathrm{cos}(\phi)$ is called a ``Jacobian factor'' because it results from a change of basis to spherical coordinates.
In this coordinate system, $\phi$~=~$\beta-90^\circ$, and so when $\beta=180^\circ$, $\phi = 90^\circ$ and the Jacobian factor is zero.
This factor explains the lack of counts at $\beta = 180^\circ$.
Importantly, the Jacobian factor has been corrected for (divided out) in all of the 1-dimensional angular distributions presented within the main text.
This is why, for example, the 1-dimensional distributions of $\beta$ and $\theta$ shown in Figs.~2G and 2H, do not tend toward zero at $\beta=180^\circ$ or $\theta=0/180^\circ$.
However, this factor is still present in all other representations of the data, such as the 2-dimensional Newton plots shown in Figs.~2A--2E, leading to a noticeable lack of counts at $\beta=180^\circ$.
\\
\section{Theoretical Methods}
\subsection{Comparing Experimental and Theoretical Trajectories} 
\indent To achieve the best possible agreement between theoretical and experimental observables, as displayed in Figs.~3A--3F, the theoretical populations on each of the nine states of D$_2$O$^{2+}$ were optimized to best reproduce the data.
This was accomplished by binning the theoretical trajectories coarsely to match the resolution of the experimental dataset, as seen in Fig.~S1, then minimizing the difference between each time-resolved observable as the theoretical populations were varied.
Specifically, we minimized the difference between Figs.~S1A and S1D, Figs.~S1B and S1E, and Figs.~S1C and S1F simultaneously. 
The results are shown in Fig.~S1G--S1I which display the 2-dimensional difference plots between theory and experiment following optimization.
The accompanying populations on each state yielded by this analysis are shown in Table~S1. \\
\begin{figure*}[ht!]
\centering
\includegraphics[width=16cm, trim={0cm 0cm 0cm 0cm}]{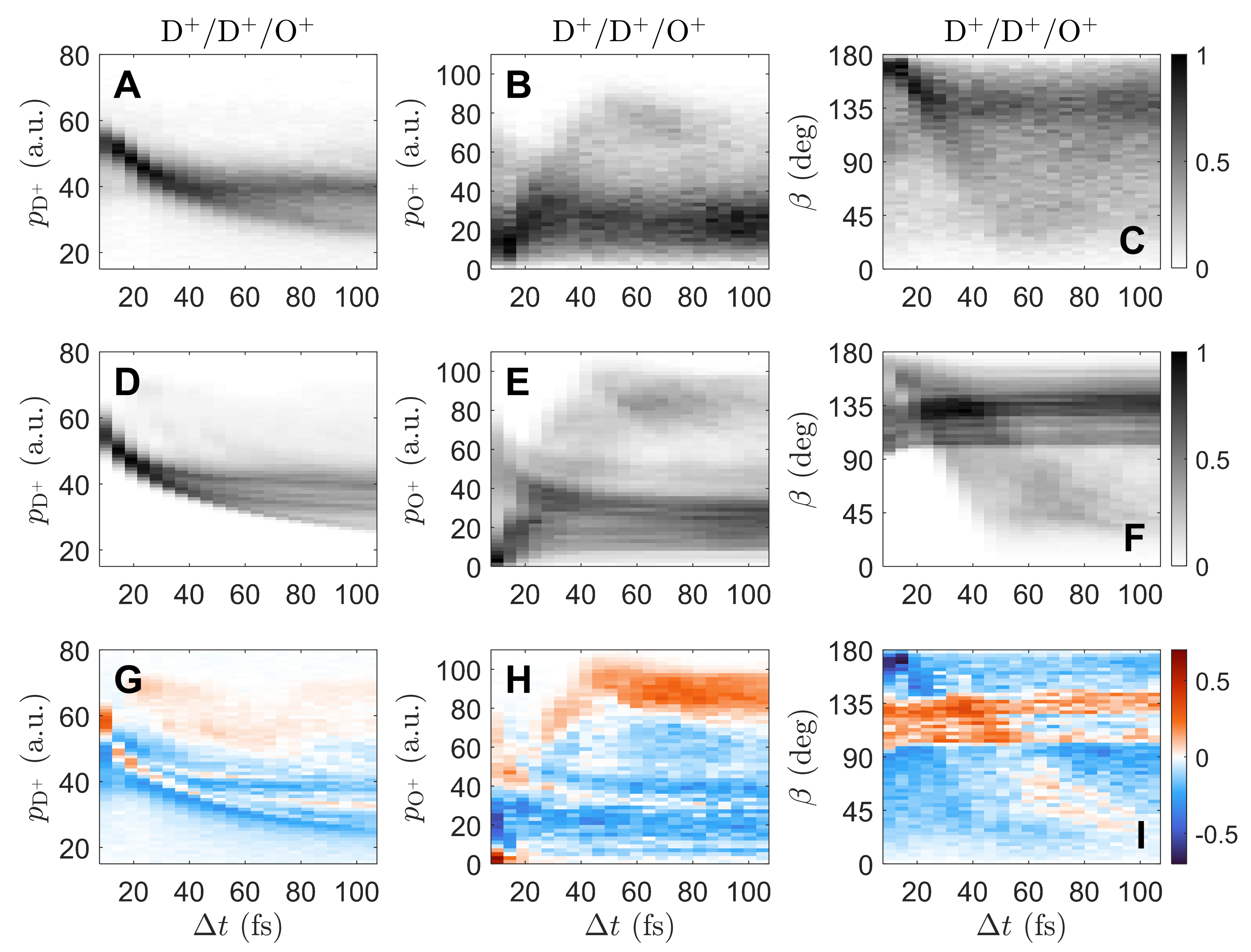}
\label{fig:DiffPlot}
\caption{ \textbf{Comparing theoretical D$_2$O$^{2+}$ dynamics to the data.} (\textbf{A--C}) Two-dimensional histograms of the three experimental observables: the magnitude of the deuteron momentum ($p_\mathrm{D^+}$), the magnitude of the oxygen-ion momentum ($p_\mathrm{O^+}$), and the momentum-frame bend angle ($\beta$), each plotted as a function of interpulse delay $\Delta t$. \mbox{(\textbf{D--F}) Two-dimensional} histograms of the same three observables as in (A--C) but for a simulated ensemble of trajectories. The normalized populations of each state are listed in Table~S1. (\textbf{G--I}) The difference between theory and experiment \mbox{(difference = theory - experiment)} for each of the observables seen in (A--C).} \vspace{0.25cm}
\end{figure*}
\indent Figs.~S1G--1I show that there is poor quantitative agreement between the data (Figs.~S1A--S1C) and simulations \mbox{(Figs.~S1D--S1F),} revealing regions in excess of 50$\%$ error.
This is expected: the EI phenomenon is highly sensitive to the geometry as well as the occupied state of the molecule.
The angular distribution in Fig.~S1C, for example, has the most dense clustering of counts at $\Delta t \sim$~18~fs and $\beta \sim$~150$^\circ$ (corresponding to $\theta_\mathrm{DOD}$~$\sim$~180$^\circ$).
The Wigner distribution of trajectories launched on any combination of the nine dication surfaces could never reproduce this clustering, because EI acts as a strong selective filter.
The closest match comes from populating the 2$^1$A$_1$ state, in which (according to Table~1) 74$\%$ of the trajectories undergo the rapid ``slingshot'' motion that is favored by EI.
The frequent occurrence of slingshot motion on the 2$^1$A$_1$ state explains why this analysis, as seen in Table~S1, yields a maximal population in this state; however, the other states are still necessary to reproduce all the features seen in the experiment.
For example, populating the higher lying states (such as $^1$B$_2$) is necessary in order to reproduce the motion in which $\beta$ unbends from 100$^\circ$ to 140$^\circ$ as $\Delta t$ progresses from 10 to 30~fs.
Ultimately, the lack of quantitative agreement between theory and experiment is a manifestation of the EI phenomenon.
Furthermore, the subset of states that are promoted to the trication can vary as a function of delay; the state population distribution extracted for early delays yields different results than those extracted for late delays, further complicating any quantitative analysis of the state populations.
For all of these reasons, the populations displayed in Table~S1 should not be considered an accurate depiction of the initial ensemble of dication states launched by the initial pulse in the pair. \\

\begin{SCtable}[2][ht!]
\centering
\begin{tabular}{ |c|c|c|c|c| } 
 \hline
 \textbf{State} & \textbf{Population} \\ 
 \hline
 $^3$B$_1$ & 0.37\\
 \hline
 1 $^1$A$_1$ & 0.49\\ 
 \hline
 $^1$B$_1$ & 0.02\\
 \hline
 $^3$A$_2$ & 0.11\\
 \hline
 $^1$A$_2$ & 0.13\\
 \hline
 2 $^1$A$_1$ & 1.00\\
 \hline
 $^3$B$_2$ & 0.04\\
 \hline
 $^1$B$_2$ & 0.44\\
 \hline
 3 $^1$A$_1$ & 0.25\\
 \hline
\end{tabular}
\caption{ \textbf{Theoretical populations of D$_2$O$^{2+}$ extracted from the data.} The normalized population within each dication state (labeled by C$_\mathrm{2v}$ symmetry) following optimization between experiment and theory. Optimization was performed to minimize the difference of the three following time-resolved observables: the magnitude of the deuteron momentum ($p_\mathrm{D^+}$), the magnitude of the oxygen-ion momentum ($p_\mathrm{O^+}$), and the momentum-frame bend angle ($\beta$). Here, the highest population is in the 2$^1$A$_1$ state.}
\end{SCtable}

\subsection{Analysis of the 3-Dimensional Enhancement Volume}
\indent Before constructing a 3-dimensional space to localize the enhancement, we first applied a strict filter in time.
To do so, we plotted the normalized yield of all D$^+$/D$^+$/O$^+$ coincidences as function of $\Delta t$ and fit this distribution to a simple Gaussian.
As seen in Fig.~S2, this normalized yield is well approximated by a Gaussian distribution at early interpulse delays.
We then defined our filter as the 6-fs window centered around the peak of this fit: (12~fs~$< \Delta t <$~24~fs).
After applying this filter to all D$^+$/D$^+$/O$^+$ coincidences, we constructed a 3-dimensional histogram in $\Delta t$, $\beta$, and $p_\mathrm{D^+}$. 
Drawing an isointensity surface at 50$\%$ of the maximum value in this 3-dimensional histogram yields the 3-dimensional enhancement volume pictured in Fig.~4A. \\
\indent As stated in the main text, we then utilized this 3-dimensional enhancement volume in order to recover the geometry associated with EI.
To do so, we propagated all 18,432 trajectories (2048 trajectories per state $\times$ 9 states) through this 3-dimensional space and assigned a weight to each trajectory per time-step (in $\Delta t$) based on the local value of the enhancement at that point within the 50$\%$ enhancement volume.
The weight assigned outside of the enhancement volume was zero.
If, for example, a trajectory passes through the global maximum of the 3-dimensional histogram, it is assigned a weight of 1 at that particular time-step.\\
\indent The ultimate result of this analysis is a list of 10,212 trajectories (8220 were eliminated entirely), each of which carries a weight that is a function of $\Delta t$.
The weighted sum of these trajectories was then used to generate the 2-dimensional histogram in $r_\mathrm{OD}$ and $\theta_\mathrm{DOD}$ that is shown in Fig.~4B. 
As there is nothing experimentally distinguishing $r_\mathrm{OD}^{(1)}$ and $r_\mathrm{OD}^{(2)}$, Fig. 4B is the average between the 2-dimensional histograms of [$r_\mathrm{OD}^{(1)}$ and $\theta_\mathrm{DOD}$] and [$r_\mathrm{OD}^{(2)}$ and $\theta_\mathrm{DOD}$].
\begin{SCfigure}[1.5][ht!]
\centering
\includegraphics[width=8cm, trim={0cm 0cm 0cm 0cm}]{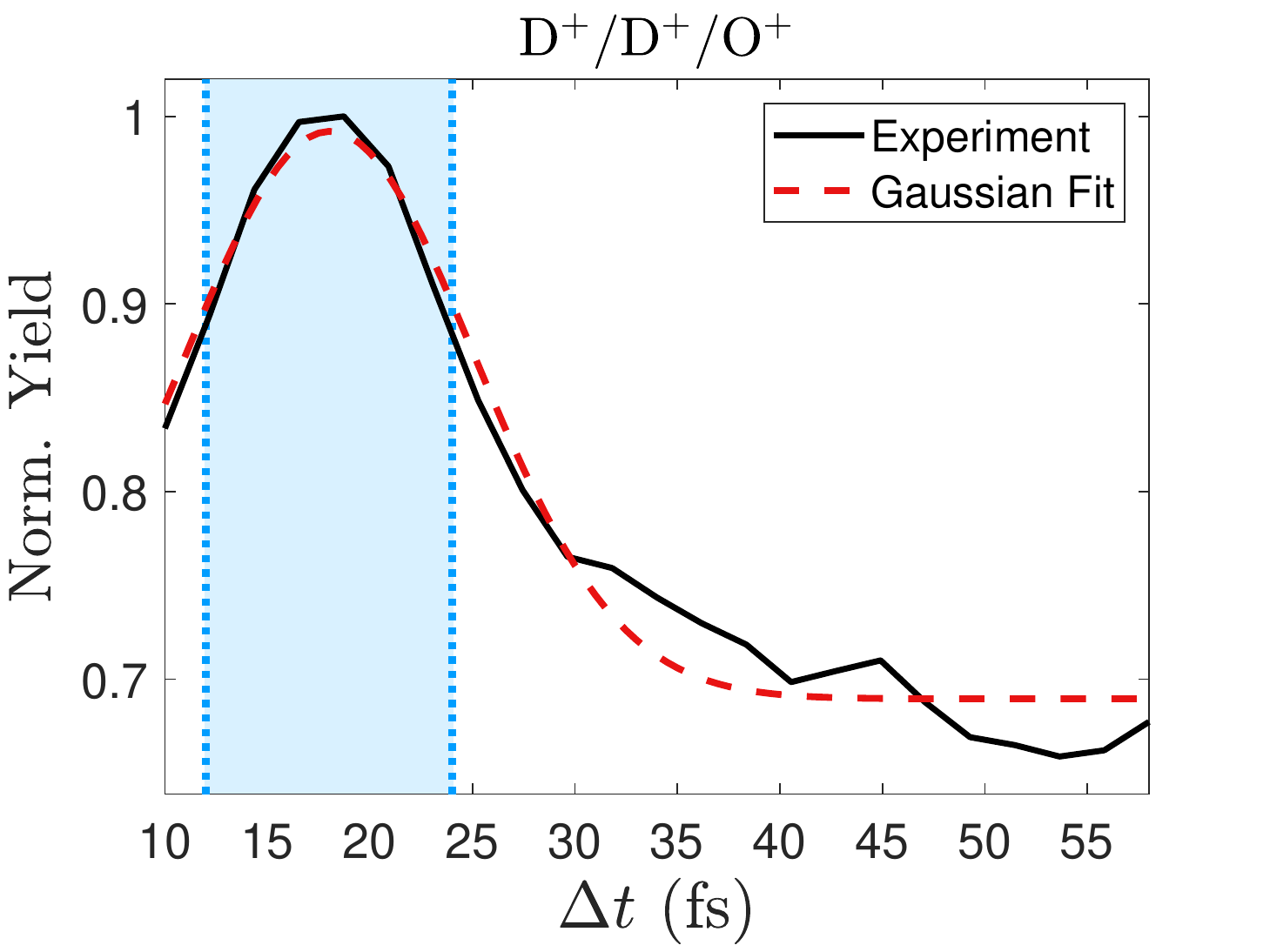}
\label{fig:Yield}
\caption{ \textbf{Delay-dependent yield of D$_2$O$^{3+}$.} The normalized yield of D$^+$/D$^+$/O$^+$ coincidences as a function of interpulse delay $\Delta t$ (in solid black) and a simple Gaussian fit to this yield (in dashed red). Highlighted in cyan is a 6-fs wide time-window centered around the interpulse delay at which the peak of the Gaussian fit occurs: $\Delta t$~=~18~fs.\\ \\ \\ \\ \\}
\end{SCfigure}


\subsection{Comparing Single-Pulse and Double-Pulse KER Distributions}
\indent To address one possible reason for the disparity in KER between the three-body dissociations following formation of D$_2$O$^{3+}$ via single pulses (where $\tau \geq$~19~fs) and pulse pairs (where $\Delta t$~=~18~fs), we make use of a simple model of CEI.
In Fig.~S3, we used the three-charge Coulomb repulsion potential (Eq.~3) to calculate the KER for the Coulomb explosion of a static and linear D$_2$O$^{3+}$ molecule as a function of symmetric stretch of the OD bond length $r_\mathrm{OD}$.
Here we assume that the DOD bend angle is 180$^\circ$ because this is a prerequisite for the EI phenomenon described in the main text.
According to this model, the peak of the KER distribution measured using 19-fs single pulses (17.5 eV) equates to an OD bond length of 1.66~$\text{\AA}$.
However, using the same model to find the equivalent OD bond length for the peak of the KER distribution measured using pulse pairs with 18-fs delay yields 2.03~$\text{\AA}$ (see Fig.~S3).
This differs significantly from the critical OD bond length (2.2~$\text{\AA}$) recovered in the main text.
The reason for this disagreement is the additional kinetic energy accumulated by propagation on the states of D$_2$O$^{2+}$.
By contrast, the simulated KER for the slingshot trajectory (the same trajectory shown in Figs.~2G--2K) takes account of this effect and correctly yields a KER of 17.5~eV for an OD bond length of 2.2~$\text{\AA}$ (see Fig.~S3).
This analysis therefore suggests that the KER found when ionizing with single pulses ($\tau \geq$~19~fs) may have a significant contribution (on the order of 1-2~eV) from the kinetic energy accumulated in the intermediate charge states (D$_2$O$^{+}$ or D$_2$O$^{2+}$).
As a result, the OD bond length for this case is likely greater than 1.66~$\text{\AA}$ and may be closer to 1.8~$\text{\AA}$.
\\
\begin{SCfigure}[1.5][ht!]
\centering
\includegraphics[width=9cm, trim={0cm 0cm 0cm 0cm}]{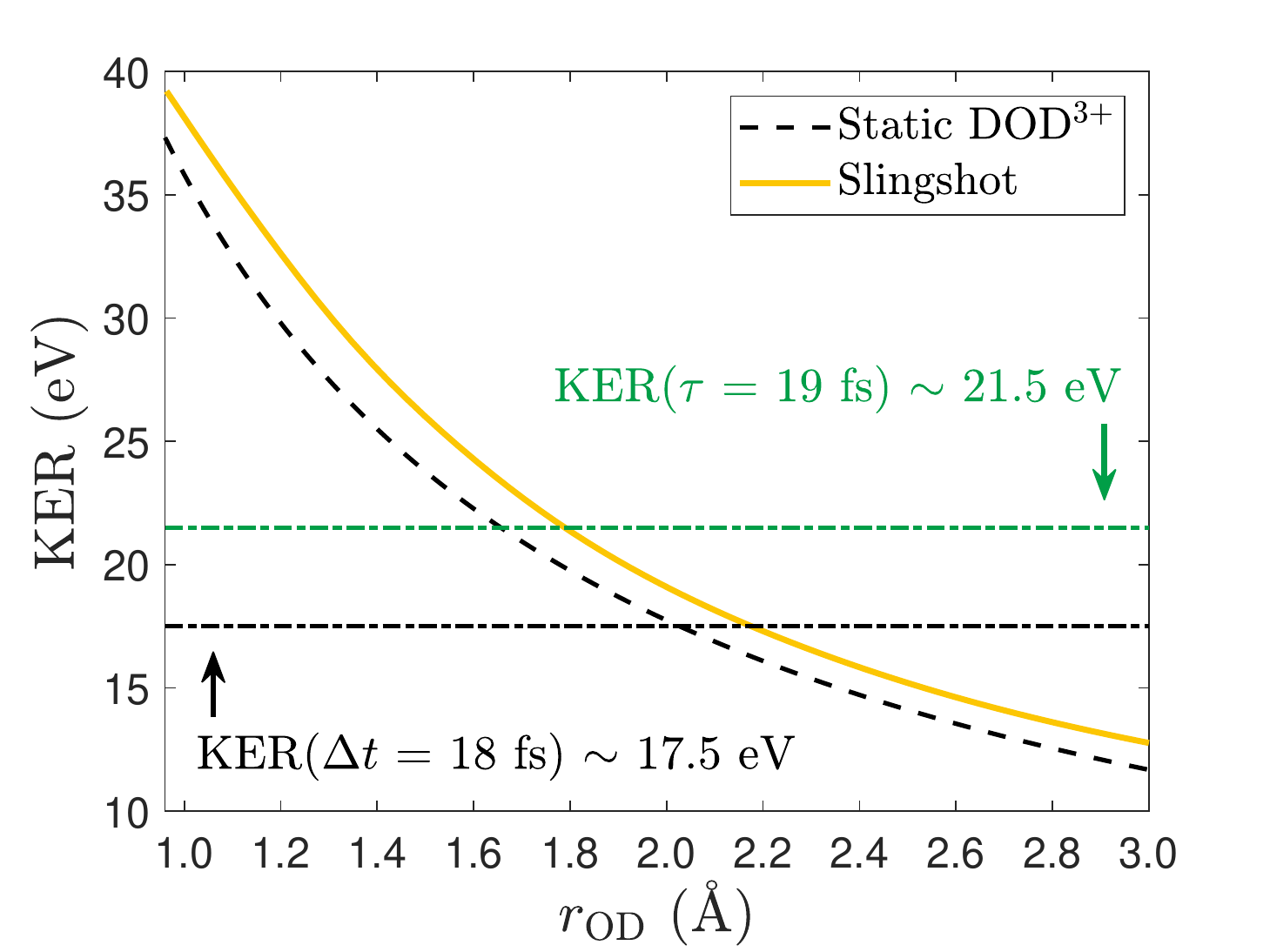}
\label{fig:KER_vs_rOD}
\caption{ \textbf{KER of D$_2$O$^{3+}$ with and without intermediate dynamics.} The total kinetic energy release (KER) as a function symmetric stretch of the OD bond ($r_\mathrm{OD}$) for the three-body Coulomb explosion of a static linear DOD molecule (dashed black line) and the slingshot trajectory first seen in Figs.~3G--3K (solid yellow line). 
Two values of KER are represented as horizontal lines: 17.5~eV and 21.5~eV.
These values correspond to the peak of the KER distribution for double pulses at a delay of $\Delta t$~=~18~fs (black dash-dotted line) and single pulses with a duration of $\tau$~=~19~fs (green dash-dotted line) respectively. \\ \\ \\ \\}
\end{SCfigure}
